\documentclass[twocolumn,10pt,pra,nofootinbib]{revtex4-1}
\pdfoutput=1
\def\mysection#1{{\bf #1.} }
\def\mysections#1{{\bf #1.} }
\usepackage{hyperref}
\usepackage{epsfig,amsmath,amssymb,verbatim,mathrsfs,array,layout,textcomp,amssymb,latexsym,slashed,graphicx,booktabs,color,mathtools,tikz,natbib}
\usepackage{cleveref}
\usepackage{diagbox}
\usepackage[flushleft]{threeparttable}
\usepackage{footnote}

\usepackage{soul}
\graphicspath{{/}}

\newcommand{\bea}{\begin{eqnarray}}
\newcommand{\eea}{\end{eqnarray} }

\newcommand{\bv}{\left(\begin{array}{c}}
\newcommand{\ev}{\end{array}\right)}
\newcommand{\bmtwo}{\left(\begin{array}{cc}}
\newcommand{\bmthree}{\left(\begin{array}{ccc}}
\newcommand{\emn}{\end{array}\right)}
\newcommand{\bmtwoc}{\left\{\begin{array}{cc}}
\newcommand{\bmthreec}{\left\{\begin{array}{ccc}}
\newcommand{\emnc}{\end{array}\right\}}
\newcommand{\ba}{\begin{array}}
\newcommand{\ea}{\end{array}}

\newcommand{\units}[1]{\mathrm{\; #1}}

\def\lsim{\mathrel{\rlap{\lower4pt\hbox{\hskip1pt$\sim$}}
     \raise1pt\hbox{$<$}}}         
\def\gsim{\mathrel{\rlap{\lower4pt\hbox{\hskip1pt$\sim$}}
     \raise1pt\hbox{$>$}}}         

\begin{document}
\font\mini=cmr10 at 0.8pt

\title{
Axion-like Relics: New Constraints from Old Comagnetometer Data
}

\author{Itay M. Bloch${}^{1}$}\email{itay.bloch.m@gmail.com }
\author{Yonit Hochberg${}^{2}$}\email{yonit.hochberg@mail.huji.ac.il}
\author{Eric Kuflik${}^{2}$}\email{eric.kuflik@mail.huji.ac.il}
\author{Tomer Volansky${}^{1}$}\email{tomerv@post.tau.ac.il}
\affiliation{${}^1$Department of Physics, Tel Aviv University, Tel Aviv, Israel}
\affiliation{${}^2$Racah Institute of Physics, Hebrew University of Jerusalem, Jerusalem 91904, Israel}

\begin{abstract}
The noble-alkali comagnetometer, developed in recent years, has been shown to be a very accurate measuring device of anomalous magnetic-like fields. An ultra-light relic axion-like particle can source  an anomalous field that permeates space, allowing for its detection by comagnetometers. Here we derive new constraints on relic axion-like particles interaction with neutrons and electrons from old comagnetometer data. We show that the decade-old experimental data 
place the most stringent terrestrial constraints to date on ultra-light axion-like particles coupled to neutrons. The constraints are comparable to those from stellar cooling, providing a complementary probe. Future planned improvements of comagnetometer measurements through altered geometry, constituent content and data analysis techniques could enhance the sensitivity to axion-like relics coupled to nucleons or electrons by many  orders of magnitude.  
\end{abstract}

\maketitle

\section{Introduction}\label{sec:intro}

Observations suggest that roughly $85\%$ of the matter content in our universe 
is in the form of Dark Matter (DM)~\cite{Ade:2013zuv,Zwicky:1933gu,Bosma:1981zz,Rubin:1985ze,Frenk:2012ph}. 
A particularly interesting class of models that may play the role of DM are  known as Axion Like Particles (ALPs). 
Originally postulated in Refs.~\cite{Peccei1977,Wilczek1978,Weinberg1978}, the axion is a pseudo-Goldstone boson which addresses the strong CP Problem~\cite{Kim:1979if,Shifman:1979if,Dine:1981rt,Zhitnitsky:1980tq,Kim:2008hd}.  ALPs, the generalization of the axion, have similar interactions  but need not address the strong CP problem.  
  Many such scenarios have been studied in the literature (see {\it e.g.} Refs.~\cite{Wilczek:1982nt,Chikashige:1980ui,Gelmini:1980re,Hook:2018dlk,Svrcek:2006yi,Arvanitaki:2009fg,Graham:2015cka,Agrawal:2018mkd,Marsh:2017hbv,Craig:2018kne,Marques-Tavares:2018cwm,Agrawal:2017eqm}) and, depending on the cosmological history, both axions and ALPs may be dark matter in some regions of their parameter space~\cite{Sikivie:2006ni,Visinelli:2009kt}.

Experiments that search for ALPs utilize their coupling to photons~\cite{Anastassopoulos:2017ftl,Zioutas:2017klh,Dafni:2015lma,Salemi:2019xgl},  gluons~\cite{Garcon:2017ixh}, electrons~\cite{Bloch:2016sjj,Hochberg:2016ajh,Terrano:2019clh,Aprile:2014eoa,Armengaud:2013rta,Redondo:2013wwa,Hochberg:2016sqx,Terrano:2015sna}, anti-protons~\cite{Smorra:2019qfx},  and protons or neutrons~\cite{Graham:2017ivz,Adelberger2007,Abel:2017rtm,Garcon:2019inh,Wu:2019exd,VasilakisThesis}. In this paper we focus on ALP couplings to electrons and neutrons, re-analyzing decade-old published data from Refs.~\cite{VasilakisThesis,KornackThesis,BrownThesis} to search for ALPs as an anomalous magnetic-like field~\cite{OldWind} that interacts with the spins of the nuclei of a helium sample, or the electrons of a potassium vapor sample.

Our limits utilize an experimental device called a helium-potassium (${}^3$He-K) comagnetometer~\cite{BrownThesis,KornackThesis,VasilakisThesis,Kornack:2004cs,Vasilakis:2008yn,Brown:2010dt,Lee:2018vaq,Romalis1}. The comagnetometer is sensitive to the difference between the magnetic fields measured by two strongly interacting magnetometers. The first  measures the magnetic field via the spin of helium-3 atoms,   which is   dominated by the spin of their neutrons. 
 The second magnetometer is sensitive to the spin of potassium atoms, which are  dominated by the spin of their outermost electron. 
The comagnetometer resonantly couples the two magnetometers, and the result is a device that is sensitive to low-frequency, $\lsim\mathcal{O}(100)\units{sec^{-1}}$, spin-dependent interactions that couple differently to neutrons and  electrons.

 The basic idea is as follows. The sensitivity of the comagnetometer is optimized at the so-called  `compensation point'. There, the response of the helium spins is tuned such that they cancel the effect of magnetic fields on the alkali (potassium) spins,  making the alkali magnetometer insensitive to regular magnetic fields. 
Anomalous magnetic fields---which couple differently to neutrons and electrons compared to regular magnetic fields---would not be canceled by the helium gas, and will have a measurable effect on the alkali. For an ALP, the ratio of its coupling to neutrons, $g_{aNN}$, to its coupling to electrons, $g_{aee}$, should generically differ from the neutron to electron gyromagnetic ratio,  and so the comagnetometer is a sensitive instrument for detecting the new magnetic-like fields that an ultra-light ALP would induce. 

As a result, the ${}^3$He-K comagnetometer can be used as a tool to measure the interactions of ALPs with neutrons and electrons. As we will show, this setup enhances the signal from ALP-neutron coupling compared to that of the ALP-electron coupling, yielding moderate sensitivity to the latter and excellent sensitivity to the former. The bounds we recast from the published data of Refs.~\cite{KornackThesis,BrownThesis,VasilakisThesis} place the strongest terrestrial constraints on the coupling of ALPs to neutrons over a broad range of masses, comparable and complementary to known astrophysical bounds.  

We note that Ref.~\cite{PospelovSlides} (as well as Ref.~\cite{Stadnik:2017mid}) suggested doing an analysis such as the one presented in this paper, and Ref.~\cite{Graham:2017ivz} (discussed in further details by Ref.~\cite{Alonso:2018dxy}) has implemented the analysis  for the case where the ALP's inverse-mass is much larger than the total measurement time, placing limits for  $m_a\lesssim 2\times 10^{-22}\units{eV}$. Our analysis lays out the machinery (distinct from that presented in Ref.~\cite{Graham:2017ivz}) needed to explore higher masses, extending the limits up to $m_a\lsim 4\times 10^{-13}\units{eV}$. 
We further discuss the near-future prospects of these experiments.

This paper is organized as followed. We begin in Section~\ref{sec:comag} by describing the  comagnetometer and its basic principle of operation. Section~\ref{sec:dyn} describes the dynamical equations of the comagnetometer. We discuss how the comagnetometer can be used to detect relic ALPs in Section.~\ref{sec:ALPs}. The data we use is presented in Sections~\ref{sec:data}, followed by our new derived limits in Section~\ref{sec:results}. We end by outlining possible improvements for future experiments in Section~\ref{sec:future} followed by a summary. The many appendices expand on the calculations and derivations performed throughout the paper. In Appendix~\ref{app:comag} we give a more complete derivation of the steady state solution of the comagnetometer.  Appendix~\ref{app:dynamics}  expands on the dynamical response of the system, and its steady state response to an oscillating signal. Appendix~\ref{app:directions} describes how the direction of the ALP wind  affects the signal, and how we treated it in our analysis while Appendix~\ref{app:nonDeterministic} presents the treatment of the implications of the stochastic nature of the ALP field.  Appendix~\ref{app:nuc} discusses two effects related to the nuclear spin structure, justifying choices we make in our analysis.  
Finally, Appendix~\ref{app:final} unites the results of all previous appendices and provides an explicit derivation of the  95\% C.L. bounds, accounting for the effects of noise. 

\section{The \texorpdfstring{${}^3$}{3}He-K Comagnetometer}\label{sec:comag}

The concept of the helium-potassium comagnetometer was originally proposed in Ref.~\cite{Romalis1} and further developed in Refs.~\cite{KornackThesis,BrownThesis,VasilakisThesis}. Below, we briefly describe the principles of its operation (for further details see Appendices~\ref{app:comag} and~\ref{app:dynamics}).

\begin{figure*}[th]
\includegraphics[width=0.18\textwidth]{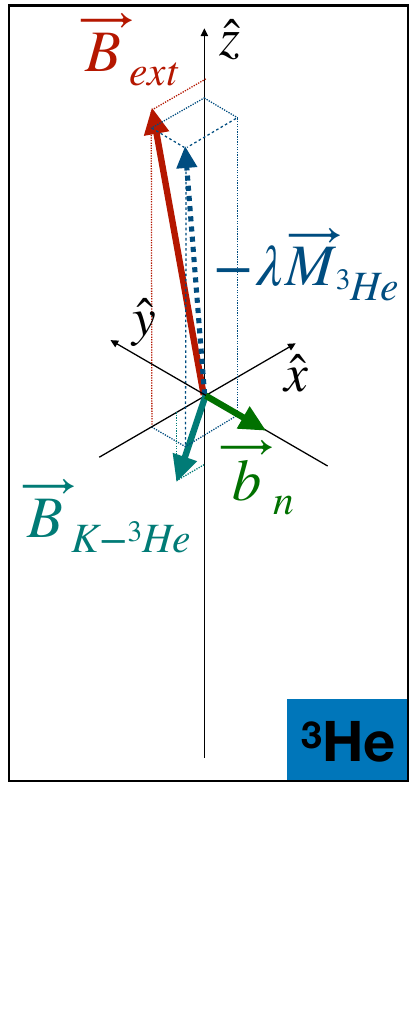} \hspace*{\fill}
\includegraphics[width=0.6\textwidth]{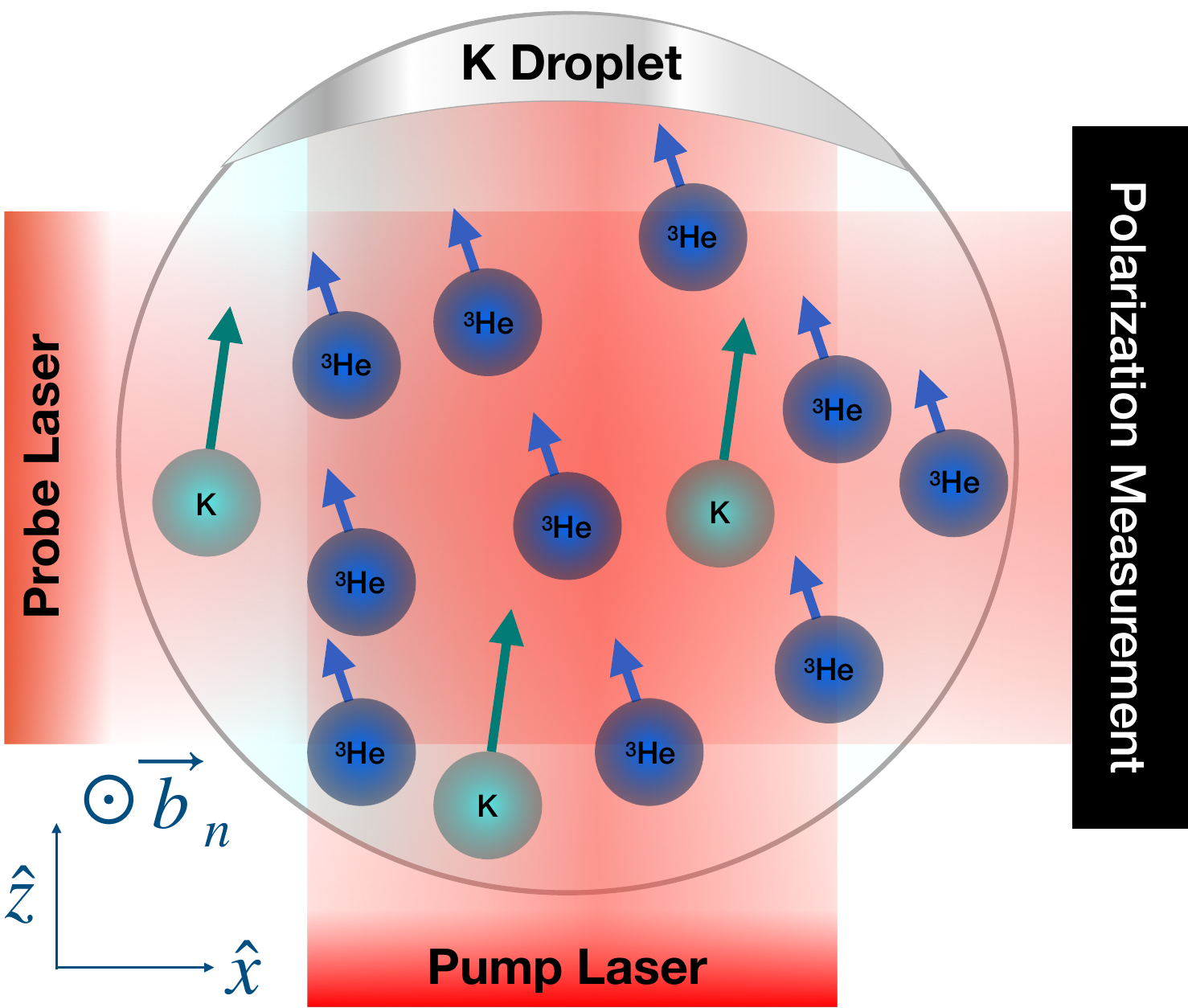} \hspace*{\fill}
\includegraphics[width=0.183\textwidth]{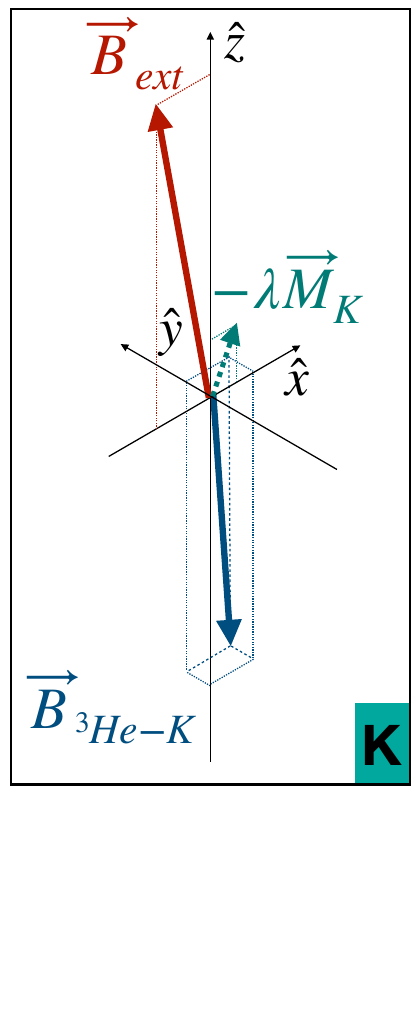}
\caption{\label{fig:comag}
{\bf Center:}  Schematic illustration of the principles of operation of the comagnetometer, including the pump laser, probe laser, polarization measurement, glass cell, K droplet  (indicated by the silver sphere),  K atoms and ${}^3$He atoms. The pump laser in the $\hat{z}$ direction polarizes the K atoms, which themselves polarize the ${}^3$He to the $\hat{z}$ direction. Measuring the outgoing probe laser beam's polarization allows one to measure the $\hat{x}$ projection of the alkali spin. In this illustration an anomalous field $\vec{b}_n$ is present ({\it e.g.} sourced by an ALP) along the $\hat{y}$ direction and affects only the ${}^3$He atoms. {\bf Side panels:} 3-dimensional axes depicting the spins of the  ${}^3$He ({\bf left}) and K ({\bf right}) and the different fields (anomalous as well as magnetic).  $\vec{B}_{{\rm K}-{}^3{\rm He}}$ ($\vec{B}_{{}^3{\rm He}-{\rm K}}$) is the magnetic field the K (${}^3$He) spins induce on the ${}^3$He (K) atoms. Both atoms are in the presence of an external magnetic field $\vec{B}_{\rm ext}$, which has a small deviation from the large, controlled $\hat{z}$ magnetic field due to magnetic noise, here assumed in the $\hat{x}$ axis. 
The overall magnetization of the K (${}^3$He) are depicted  by the dotted vectors and marked as
$-\lambda \vec{M}_{K}$ ($-\lambda \vec{M}_{{}^3 He}$).
The tuning of the $\hat{z}$ component of $\vec{B}_{\rm ext}$ to what is called the compensation point, ensures that the effect of the ${}^3$He's magnetization on the K spins, $\vec{B}_{{}^3{\rm He}}$, has a projection in the $\hat{x}$ axis which exactly cancels the effects of $\vec B_{\rm ext}$ on K. The rotation induced by $\vec{b}_{n}$ on the ${}^3$He induces transverse polarization in the perpendicular direction on the K spin. This implies the comagnetometer has sensitivity to anomalous fields, while it is insensitive to regular magnetic noise. See main text for further details.
 }
\end{figure*}

The ${}^3$He-K comagnetometer is depicted in Fig.~\ref{fig:comag}. It is a hybrid of two magnetometers that occupy the same space and interact with each other. The setup typically includes a spherical glass cell containing potassium~(K) vapor and a highly pressurized  helium-3 gas (${}^3$He). The glass cell is illuminated by two laser beams, referred to as the `pump' and `probe'. The pump beam is used to initialize the comagnetometer by polarizing the potassium atoms to its direction, while the probe  measures  the spin of the potassium atoms. The glass cell is surrounded by magnetic coils, which are themselves surrounded by magnetic shields, so that the magnetic field inside the cell remains under control to a high degree. The density of the potassium vapor is determined by the temperature of the cell, which is controlled using an oven.

\mysection{The alkaline K magnetometer}
The spins of the potassium magnetometer are initialized to a certain  direction, $\bf\hat z$, via the pump beam. Further stabilization of the polarization in this direction is achieved through the placement of a magnetic field aligned in the $\bf\hat z$-direction.  Such a magnetic field has two crucial additional roles to be discussed below: (i) it is used for mitigating magnetic noises in the ${}^3$He system and (ii) by tuning this field to a specific value one may strongly couple the two magnetometers to one another.   
A weak transverse magnetic or anomalous field, that changes slower than the decay rate of the alkali's transverse polarization (induced mostly by the pump), adiabatically tilts the spins and induces a measurable change in the direction of the alkali's polarization. Since the  alkali would only partially be able to follow fields that change too fast, its sensitivity is reduced when the typical time scale for changes in the magnetic fields is shorter than the inverse decay rate. 
The probe beam measures the projection of this polarization along its direction, while minimally affecting the alkali spins. The resulting magnetometer is sensitive to fields perpendicular to both the pump and the probe beams.\footnote{By using two probe beams, one could in principle measure the magnetic fields in the complete plane perpendicular to the pump. While this idea is not implemented in the experiments that we analyze in this work, the comagnetometer of Ref.~\cite{BrownThesis} was rotated every few seconds to achieve sensitivity to two directions in a similar manner.}

\mysection{Dynamics of helium-3 atoms}
Helium-3 is a spin-$\frac{1}{2}$ atom with its two electrons in the singlet state.  Consequently, its spin originates entirely from the nucleus. Using the pump and probe beams at wavelength $770\; {\rm nm}$, they have  practically no interactions with the nuclear energy levels associated with the helium-3 spins. 

The helium-3  dynamics 
benefit from  two important effects that stem from their  {\it spin-conserving} collisions with the alkali metal.  First, these collisions  polarize the helium-3 gas, operating as an effective pumping force that generates a macroscopic helium-3 magnetization and acts to (slowly) decay any spin component that is not aligned with the alkali polarization along the $\bf\hat z$ direction. Second, the collisions induce mutual effective  magnetic fields.  The magnetic field induced by the alkali is  significantly smaller than the external magnetic field in the $\bf\hat z$ direction, however it plays a crucial role for the dynamics of the helium-3 spins in the transverse directions, as discussed below. 

The primary goal is to measure an anomalous field transverse to the $\bf\hat z$ direction, 
 which oscillates slowly in time (much like an ultra-light axion).   To do so, timescales play an important role.  For simplicity, it is easier to think of the anomalous field as though it only interacts with either electrons or neutrons, and correspondingly affects only the potassium or the helium. As mentioned above, the response of the alkali is damped when the field oscillates much faster than the alkali's decay rate. 
 In a generic situation, the helium-3 decay rate is small, or equivalently,  the lifetime of its transverse nuclear spin excitations is very long.  Consequently, if an  anomalous field interacting only with neutrons is oscillating much faster than the lifetime, its oscillations will effectively average out before helium-3 spins have time to follow it by decaying to the direction of the net-magnetic field.
To solve this problem (as well as to probe the helium-3 spin), the system must be brought into a resonance,  which significantly shortens the transverse lifetime of the helium-3 spin.   We now discuss the method to achieve this. 

\mysection{Interactions of the two magnetometers}
With the two magnetometers placed in the same glass-cell, the system exhibits two modes, one that is mostly aligned with the short-lived  alkali metal, and the other much longer-lived mode that is mostly aligned with the spin of the noble gas.   The interactions between the two gases induce an effective coupling that triggers both the pumping effect in the helium-3 and mutual effective magnetic fields.\footnote{Note that the effective pumping of the alkali due to the presence of the helium is negligible compared to the direct pumping from the pump beam.  Conversely, the source of the helium polarization is non other than the pumping achieved by the presence of the alkali.}

The mixing, however, is a priori insufficient to significantly affect the lifetime of the helium-3 (of order a few hours; see $R_{\rm He}$ in table~\ref{table:raw}), unless the two modes are in resonance. Since the pump and external magnetic field are both aligned with the $\hat{\bf z}$ direction, the noise in the pumping rate and in the $B_z$ amplitude would dominate over any new anomalous field in the $\hat{\bf z}$ direction. Therefore, sensitive measurements cannot be implemented in the $\hat{\bf z}$ direction, and one only measures the transverse spins.

By tuning the magnetic field in the $\hat{\bf z}$ direction, one can tune the energy splitting due to $\hat{\bf z}$ magnetic fields in the two spin species to be identical, putting the two magnetometers in resonance. At this point, the two previously separable magnetometers become mixed---allowing sensitivity to the nuclear spins through the measurement of the alkali spins. Moreover, the lifetimes become similar, and in particular, the effective lifetime of the helium-3 is reduced by orders of magnitude compared to the non-resonant mode, of order $\sim100 \units{msec}$.
 
 A very important effect happens close to the resonance regime, which significantly enhances the comagnetometer sensitivity. Under steady state conditions, the nuclear polarization of the helium-3 can be made to follow external magnetic fields, thus canceling the net magnetic fields felt by the alkali (in the transverse directions). This specific choice of magnetic field is called  the {\it  compensation point}. It is usually ${\cal O}(1\%)$ away from the resonance point, thus reaping most of the sought-after benefits of the latter point as well. At the compensation point, the alkali spins---which interact with the total external and nuclear magnetic fields---feel a vanishing overall magnetic force. Consequently, the comagnetometer cancels out regular magnetic fields, leaving excellent sensitivity to anomalous ones. 

We can now expand on the schematic depiction of the comagnetometer of  Fig.~\ref{fig:comag}. In the center panel, the large circle represents the glass cell which houses a pressured ${}^3 {\rm He}$ gas, as well as a liquid silvery droplet that generates a vapor of K atoms. The probe laser passes through the glass cell, and its linear polarization is modified by the alkali spins, allowing for the 
projection of the potassium spin along the direction of the probe beam  propagation ($\equiv \hat{\bf x}$) to be measured. The pump laser is circularly polarized and interacts with the alkali atoms, giving them a macroscopic polarization in the direction of the pump beam propagation ($\equiv \hat{\bf z}$). This macroscopic polarization is passed (to some degree) to the ${}^3$He atoms, giving them a macroscopic magnetization in the pump beam's direction ($\hat{\bf z}$) as well. 
The left drawing of vectors represents the fields operating on the ${}^3{\rm He}$ atoms, with $\bf{b}_n$  the {\it anomalous field} interacting with the neutrons, 
$B_{{\rm K}-{}^3{\rm He}}$  the magnetic field the alkali induces on the ${}^3 {\rm He}$, and  ${\bf B}_{\rm ext}$  the external magnetic field. 
Note that ${\bf B}_{\rm ext}$ is not precisely along the $\hat{\bf z}$ direction due to possible experimental noise.  We have chosen to depict only ${\bf b}_n$ (and not ${\bf b}_e$ as well), in order to simplify the illustration. The right drawing of vectors represents the fields operating on the ${\rm K}$ atoms, with $B_{{}^3{\rm He}-{}^3{\rm K}}$  the magnetic field the ${}^3$He atoms induce on the K atoms. $\lambda \vec{M}_K=\vec{B}_{{K}-{}^3He}$ ($=2\lambda \mu_{\rm K}{\bf S}_{\rm K}$ in later equations) and $\lambda \vec{M}_{{}^3{ He}}=\vec{B}_{{{}^3{\rm He}}-{\rm K}}$($=2\lambda \mu_{\rm He}{\bf S}_{\rm He}$ in later equations) represent the effective magnetization of the alkali as felt by the noble gas and of the noble gas as felt by the alkali, respectively. Note that the vector $-\lambda \vec{M}_{K}$ ($-\lambda \vec{M}_{{}^3 He}$) is proportional to the direction of the spins of the K (${}^3{\rm He}$) atoms with a positive proportionality scale. $-\lambda \vec{M}_{K}$ ($-\lambda \vec{M}_{{}^3 He}$) therefore is not a field that is felt by K (${}^3{\rm He}$) spins, rather it is the direction of the K (${}^3{\rm He}$) spins. 

The compensation point is illustrated in Fig.~\ref{fig:comag} by the 3-dimensional axes showing $(\lambda \vec{M}_{\rm K}+\vec{B}_{\rm ext})\cdot \hat{\bf x}=0$, since the noble gas exactly cancels the transverse component of the external magnetic field (which in Fig.~\ref{fig:comag} is directed along the $\hat{\bf x}$ axis). On the other hand, the $\hat{\bf y}$ projection of noble spins is  non-vanishing and proportional to the nuclear anomalous field (which in Fig.~\ref{fig:comag} is along the $\hat{\bf y}$ axis). As a consequence, the $\hat{\bf y}$ anomalous field  induces, through the nuclear spins, a measurable tilt in the alkali spins along the  $\hat{\bf x}$ axis.

\section{Dynamical Equations}\label{sec:dyn}
\renewcommand{\arraystretch}{1.3}
\begin{table}[]
\begin{tabular}{|c||c|c|c|}
\hline Variable	& Ref. \cite{VasilakisThesis}	& Ref.\cite{KornackThesis}	& Ref.\cite{BrownThesis}	\\ \hline\hline
$\gamma_{e}\units{[sec^{-1}/\mu G]}$			& \multicolumn{3}{c|}{$-18$}			\\ \hline
$\gamma_n\units{[sec^{-1}/mG]}$		& \multicolumn{3}{c|}{$-18$} 			\\ \hline
$\gamma_{\rm He}\units{[sec^{-1}/mG]}$		& \multicolumn{3}{c|}{$-20$} 			\\ \hline
$q$						& $5.2$		& $5.0$		& 5.2						\\ \hline
$R_e\units{[sec^{-1}]}$					& $330$	& $400$			& $350$		\\ \hline
$R_{\rm He}\units{[10^{-6}\ sec^{-1}]}$			& $24$	& $1000$		& $200$	\\ \hline
$R_{\rm pu}\units{[sec^{-1}}]$			& $170$	& $180$ & $170$						\\ \hline
$R_{\rm pu}^{\rm eff}\units{[10^{-6}\ sec^{-1}]}$	& $1.8$	& $15$		& $4$					\\ \hline
$S_{\rm K}^{z}$			& $0.25$	& 0.27 & 0.25						\\ \hline
$S_{\rm He}^z$			& $0.017$	& $0.0046$					& $0.01$				\\ \hline
$B_c\units{[mG]}$					& $5.3$ 	& $1.6$			& $2.6$						\\ \hline
$2\lambda\mu_{\rm K}\units{[mG]}$	& $-0.028$	& $-0.02$		& $-0.056$	\\ \hline
$2\lambda\mu_{\rm He}\units{[mG]}$	& $-310$	&  $-340$			& $-260$					\\ \hline
\end{tabular}
\caption {Values of important constants in Eqs.~\eqref{eq:Sdot}-\eqref{eq:Kdot} (to two digits precision) from Refs.~\cite{KornackThesis,VasilakisThesis,BrownThesis}. As in Ref.~\cite{KornackThesis}, we use directly measured values for observables, when available.}
\label{table:raw}
\end{table}

Much of the dynamics of the comagnetometers described above can be captured by the coupled time-evolution equations for the helium-3 nuclear spin vector, ${\bf S}_{\rm He}$, and the alkali spin vector, ${\bf S}_{\rm K}$ (for further details see Appendix~\ref{app:comag}):

\bea
\dot{\bf S}_{\rm K}&=&\frac{\gamma_e}{q}\left({\bf B}+\frac{ {\bf b}_{e}}{\gamma_e}+2\lambda \mu_{\rm He} {\bf S}_{\rm He} \right)\times {\bf S}_{\rm K}\nonumber\\
&&-\frac{1}{q} R_e{\bf S}_{\rm K}+\frac{1}{2q}R_{\rm pu} {\bf s}_{\rm pu}\,,\label{eq:Sdot}\\
{\bf \dot{S}}_{\rm He}&=&\gamma_{\rm He} ({\bf B}+\frac{{\bf b}_{n}}{\gamma_{n}}+2\lambda \mu_{\rm K} {\bf S}_{\rm K})\times{\bf S}_{\rm He}\nonumber\\
&&-R_{\rm He}{\bf S}_{\rm He}+R_{\rm pu}^{\rm eff}({\bf S}_{\rm K}-{\bf S}_{\rm He})\label{eq:Kdot}\,.
\eea
The typical sizes for the different variables are shown in Table~\ref{table:raw}. The first line in each of the equations describes the action of the effective total field on the corresponding spins.  Here ${\bf B}$ is the total external magnetic field, namely the controlled magnetic field in the $\bf\hat{z}$ direction, together with any magnetic noise penetrating the magnetic shielding or generated by thermal noise in the shield.  ${\bf b}_{n}$ (${\bf b}_{e}$) is an anomalous field\footnote{Note that our definition of ${\bf b}_n$ (${\bf b}_e$) differs from that of Refs.~\cite{VasilakisThesis,BrownThesis,KornackThesis} by a factor of $\gamma_{n}$ ($\gamma_e$).}  that interacts with the neutrons (electrons). $\mu_{\rm K}$ and $\mu_{\rm He}$ are the spin-normalized alkali and noble gas magnetizations respectively, while the factor $\lambda\simeq 50 $ is related to the cross section of a nuclear-alkali collision, and depends upon the overlap of the alkali and nuclear wave-functions during a collision\footnote{In a general cell geometry, an additional classical magnetic dipole-dipole term exists that modifies this $\lambda$, however such a term averages to zero in a spherical cell. See Ref.~\cite{Romalis:2014bna} for more details.}.  Under typical conditions, $|\mu_{\rm K}{\bf S}_{\rm K}|\ll |\mu_{\rm He} {\bf S}_{\rm He}|$. $\gamma_e$  ($\gamma_n$) is the gyromagnetic ratio of a free electron (neutron), while $q$ is called the `slowing down factor' that arises from integrating over the spin-3/2 degrees of freedom of the potassium nucleus, and is a dimensionless constant of order ${\cal O}(4-6)$, depending on the precise experimental setup. Finally, $\gamma_{\rm He}$ is the ${}^3$He gyromagnetic ratio. 

The decays of the spins are described by the first term of the second line in each of the equations.  $R_e$ and $R_{\rm He}$ are the decay rates  of the electron and ${}^3$He spins respectively. Finally the effect of the external pump for the alkali and the effective pump due to the spin-exchange interactions for the helium-3 are described by the last terms. ${\bf s}_{\rm pu}$ is the spin of the circularly polarized pump beam, ${\bf s}_{\rm pu}=\hat{\bf{z}}$, 
while $R_{\rm pu}$ and $R_{\rm pu}^{\rm eff}$ are the external and effective pumping rates respectively. As can be seen, the effective pump drives the helium-3 spin to align with that of the alkali. Note that the probe beam---which can be thought of as the ability to measure the alkali's spin projection on the direction of the probe's propagation---does not appear in the above equations, as it has negligible effect on the dynamics of the potassium spins and none at all on the ${}^3$He atoms.
 
The system can be understood in a simple manner under the assumption of a steady-state equilibrium, $\dot{\bf S}_{\rm K, He} = 0$. The pumping terms dominate the steady state solution of the $\bf\hat z$ projections, greatly impairing the sensitivity to all fields in that direction. Conversely, significant sensitivity can be achieved in the perpendicular directions when in the so-called compensation point. As we show in Appendix~\ref{app:comag}, in the absence of an anomalous field, ${\bf b}_{e,n}=0$, the transverse spin polarization of the alkali gas can be made to vanish (even for finite ${\bf B}_{\perp}$), ${\bf S}^\perp_{\rm K}=0$, by tuning the $\bf\hat{z}$ component of the external magnetic field to be
\begin{equation}
B_z = B_c\equiv -2\lambda \left(\mu_{\rm He} S_{\rm He}^z + \mu_{\rm K} S_{\rm K}^z\right)\,,
\end{equation}
 where $B_c$ is the  {compensation point} of the magnetic field.   Correspondingly, one often defines the {\it compensation frequency}, given by $\omega_c \equiv \gamma_{\rm He} B_c$, which is usually the typical time-scale that characterizes the compensation point. At this point, the alkali gas feels no (non-anomalous) external magnetic fields in the perpendicular direction. We stress that this ability to cancel external magnetic fields is achieved by only tuning the controlled magnetic field along the ${\bf\hat z}$ direction, allowing to cancel any additional noise in the system.  As a consequence of the compensation point, the sensitivity to anomalous fields acting on the neutrons is maximized and is found to be (see Appendix~\ref{app:comag}).
 \begin{equation}\label{eq:sigperp}
{\bf S}_{\rm K}^\perp = -\frac{1}{R_e}\left(\frac{\gamma_e}{\gamma_n}{\bf b}^\perp_n - {\bf b}^\perp_e\right)\times (S_{\rm K}^z \hat{\bf z}).
\end{equation}
The above shows an enhanced sensitivity to ${\bf b}^\perp_n$ due to the large numerical coefficient, $\gamma_e/\gamma_n \simeq {\cal O}(1000)$. The compensation point occurs within the resonance regime where the decay rate of the helium-3 is highly enhanced.

\section{Measuring ALPs with the comagnetometer}
\label{sec:ALPs}
Having established the basic concept of comagnetometers, we move to discuss their sensitivity to new physics. In particular, we focus on the ALP Langrangian terms~\cite{OldWind}, 
\begin{equation}
\label{eq:lterm}
\mathcal{L}=-g_{aNN}\partial_\mu a \bar{N}\gamma^5 \gamma^\mu N-g_{aee}\partial_\mu a \bar{e}\gamma^5 \gamma^\mu e\,,
\end{equation}
where $a$, $N$ and $e$ are the ALP, neutron and electron fields, respectively while $g_{aNN}$ and $g_{aee}$ are the ALP-neutron and ALP-electron couplings. 

The non-relativistic limit of the above results with   spin-dependent interactions  that are analogous to the interactions of magnetic fields and spins in the SM.  In analogy to magnetic fields, we define an ALP-induced field ${\bf b}$, which couples to the alkali's spin (dominated by its electronic configuration) and the helium-3 spin (governed by the spin of its neutron) with the following Hamiltonian:
\begin{equation}
\label{eq:anomalous}
H=\frac{g_{aee}}{q} {\bf b} \cdot {\bf S}_{\rm K}+\frac{\gamma_{\rm He}}{\gamma_n}g_{aNN} {\bf b} \cdot {\bf S}_{\rm He}\,.
\end{equation}
As mentioned previously, such a field is called anomalous if the ratio of the above couplings $g_{aee}/g_{aNN}$ does not match that in the Standard Model (SM), $\gamma_e/\gamma_n$.  Microscopically this is the case for any force mediator that couples differently to electrons and neutrons. (For this reason,  comagnetometers are not sensitive to relic dark photons, which would couple with the same ratio as the SM photon.)  

As described above, comagnetometers are excellent detectors of such anomalous fields, with best sensitivity demonstrated for anomalous couplings to nucleons. Refs.~\cite{KornackThesis,BrownThesis} performed a thorough search, with the results mostly interpreted in the context of anomalous fields sourced by Lorentz violation, thereby considering only time-independent anomalous fields, ${\bf b}_n \equiv g_{aNN} {\bf b}$. In this work, we show that the same data can be used to place constraints on anomalous fields that are sourced by the presence of relic ALPs that induce an effective {\it time-dependent} ${\bf b}_n$, oscillating at a frequency related to their mass, $m_a$. Bounds can similarly be placed for ALP-electron anomalous fields, ${\bf b}_e\equiv g_{aee}{\bf b}$, though these are somewhat weaker. 

When a spin-1/2 neutron, $N$, is in the presence of a coherent ALP field, $a$, assuming both are non-relativistic, the Hamiltonian of their interaction is given by~\cite{OldWind,Garcon:2017ixh,Stadnik:2013raa}
\begin{equation}
\label{CasperEQ}
H_{aNN}=g_{aNN}\sqrt{2\rho_{a}} \cdot \cos(E_a t+\theta_0) \left( {\bf {v}}\cdot {\bf \sigma}_N\right).
\end{equation}
where $\rho_{a}$ is the energy density of the ALPs at the vicinity of the neutron. $\theta_0$ is some random initial relative phase. ${\bf \sigma}_N$ is the spin of the neutron, and $E_a$ is the energy of the ALP, which for a non-relativistic particle is roughly its mass, $E_a \simeq m_a$. The relative velocity between the neutron and the ALP field is ${\bf v}$, and for DM ALPs we have on average $|{\bf v}|\sim 7.7\times 10^{-4} $ in natural units. The Hamiltonian of the interaction between an ALP and electrons is similar, with the replacements $g_{aNN}\rightarrow g_{aee},{\bf \sigma}_N\rightarrow {\bf\sigma}_e$.\footnote{One might wonder whether the wavefunctions of the bounded neutron or electron could introduce non-trivial effects, such as  modifying the relative velocity by order $\sim \alpha_{EM}$. However, as the ALP field is nearly constant over the size of an atom in the mass range we consider, when integrating over the hamiltonian density to reach the interaction Hamiltonian, such effects are averaged out.} As is evident, relic ALPs would act as an anomalous field: they couple to the spin of the particles with an oscillating strength, and an effective anomalous field, 
\begin{equation}
\label{eq:ALPsbfield}
{\bf b}_{n} = g_{aNN}\sqrt{2\rho_a}\cos(E_a t+\theta_0) {\bf v}\,,
\end{equation}
with a corresponding equation for electrons.

\section{Data}
\label{sec:data}

\begin{figure}
	\includegraphics[width=0.48\textwidth]{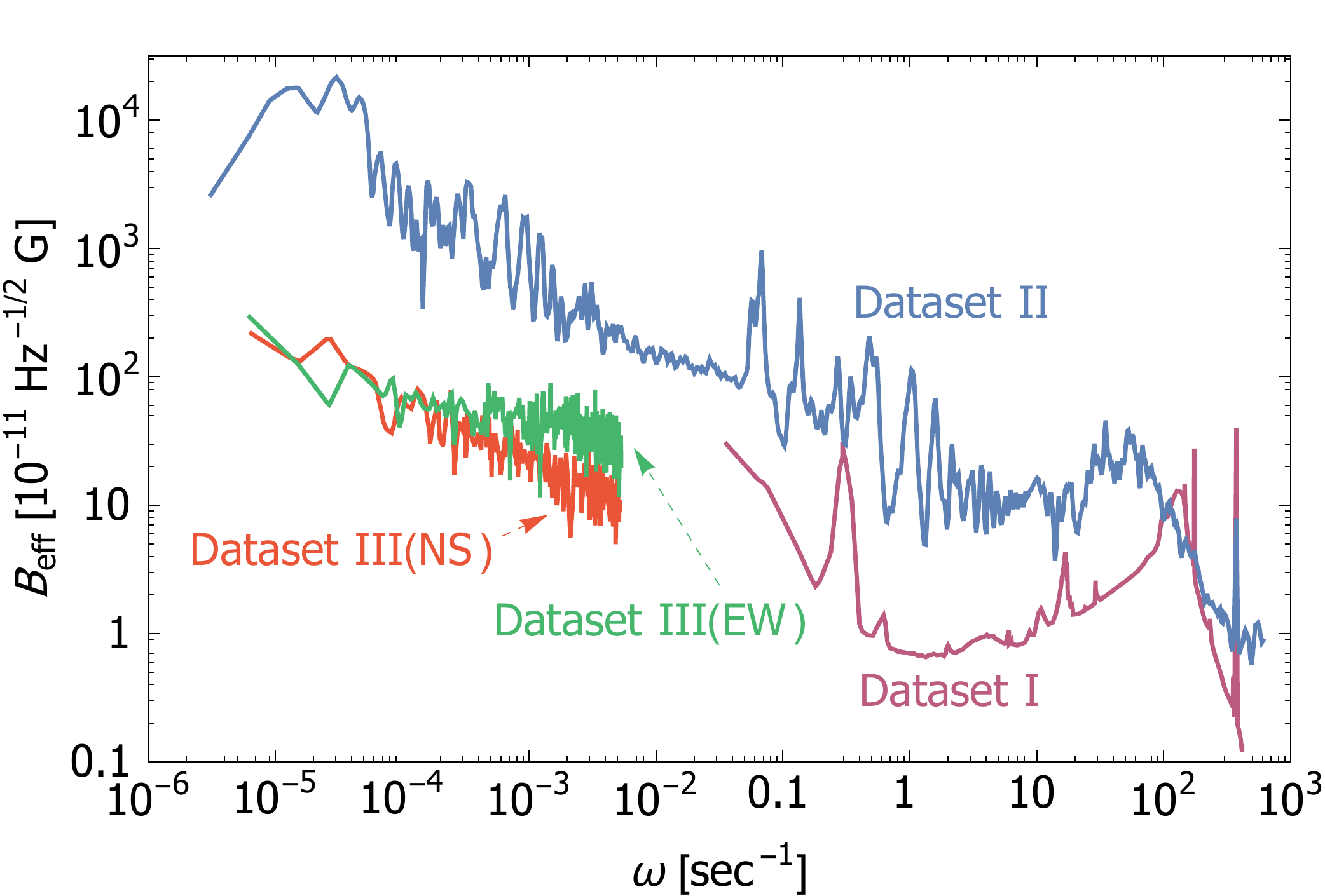}
	\caption{\label{fig:data}
The spectral noise density as a function of the angular frequency collected from three experiments and used to derive the limits in this work. Dataset I (purple) is taken from Figure 5.3 of Ref.~\cite{VasilakisThesis}, dataset II (blue) is taken from Figure 4.17 of Ref.~\cite{KornackThesis}, and dataset III (orange and green) are taken from Figure 5.11 of Ref.~\cite{BrownThesis}. Note that we present the data as a function of the angular frequency $\omega=2\pi f$, instead of as a function of frequency itself $f$.}
	\end{figure}

 In this work we analyze existing data and show how it can be used to place new bounds on ALP couplings. The data comes from three experiments, each measuring the spins of potassium atoms in a ${}^3$He-K comagnetometer for a period of several days. The data, reproduced in Fig.~\ref{fig:data}, is given in the form of the magnetic field spectral noise density, 
\begin{equation}
\label{eq:RomSig}
A(\omega)=\frac{1}{\gamma_{n}\sqrt{t_{\rm tot}}} \left|\int_0^{t_{\rm tot}}dt \left( {\bf b}_n\cdot \hat{\sigma}\right)  e^{i \omega t} \right|\,.
\end{equation}
Here $A(\omega)$ is the amplitude of the signal in units of magnetic field per square root bandwidth (where bandwidth is measured in units of ${\rm Hz}$), $\omega$ is the frequency at which the signal is measured, $t_{\rm tot}$ is the total measurement time, $\hat{\sigma}$ is the direction for which the measurement is sensitive, and ${\bf b}_n$ is the neutron anomalous field. A similar equation can be written if one assumes a signal from an electron anomalous field (${\bf b}_n\to \gamma_n{\bf b}_e/\gamma_e$).

The details of the three data sets we use are as follows: 
\begin{enumerate}

\item {\bf Dataset I:} Vasilakis et al., Ref.~\cite{VasilakisThesis} (some of the data is only shown in a plot in Ref.~\cite{BrownThesis}), performed a search for a long-range spin-dependent interactions using a comagnetometer, over a total integration time of 36.2~days. In this experiment, relic ALPs would appear as a background field, and thus the noise spectrum they provide can be used to constrain such a relic. 

The available data, which is depicted by the purple curve of Fig.~\ref{fig:data}, presents the measured noise spectrum for the entire experiment, for frequencies in the range of $0.04\units{sec^{-1}}\lesssim\omega\lesssim400\units{sec^{-1}}$. The data above $315\units{sec^{-1}}$ is filtered and therefore cannot be used to derive bounds. 

The experiment was split into 7 separate runs, each testing different configurations which affect the long-range spin-dependent interaction search, but do not affect the sensitivity to relic ALPs. However, since the different measurements have been summed incoherently, and with long breaks between them, there are non-trivial effects, discussed in further details in Appendices~\ref{app:nonDeterministic} and~\ref{app:final}. Throughout the experiment, the sensitive direction of the comagnetometer was aligned with the radial direction of the earth. This dataset will be used to derive new limits on ALP-neutron and ALP-electron couplings for masses in the range $2.4\times 10^{-17} \units{eV}\lsim m_a\lsim 2\times 10^{-13}\units{eV}$.

\item {\bf Dataset II:} The second dataset was presented by Kornack et al. in Ref.~\cite{KornackThesis} and is available only for a measurement period of $6~{\rm days}$, for frequencies $3\times 10^{-6}\units{sec^{-1}}\lesssim\omega\lesssim 600\units{sec^{-1}}$. Throughout, the sensitive direction of the comagnetometer was aligned with the radial direction of the earth.

The data is presented by the blue curve of Fig.~\ref{fig:data}. This dataset is the oldest of those we use and is noisier over most of its covered frequencies. Additionally, its resolution over the frequency range which is uncovered by other measurements is too poor to detect daily modulation, which---combined with its single measurement source---further suppresses its reach (see Appendix~\ref{app:final} for further details). This data is used to cast new bounds in mass regions not covered by the other datasets, for $3\times 10^{-18} \units{eV}\lsim m_a\lsim 4\times 10^{-17} \units{eV}$ and $2\times 10^{-13} \units{eV}\lsim m_a\lsim 4\times 10^{-13} \units{eV}$. (While this dataset could be used to cast limits on arbitrarily low ALP masses, it is non-competitive with results derived from dataset III below and so we do not pursue this further.)

\item {\bf Dataset III:} The third dataset was presented by Brown et al.~in Ref.~\cite{BrownThesis} and is available only for their longest uninterrupted measurement, which lasted $21.81$ days out of $143$ days of total run time.
Every $7-10$ seconds, the sensitive direction of the comagnetometer was rotated by $90^\circ$. The sensitive directions of the available measurement in this case are therefore both north-south and east-west. 

The measured noise spectral density is depicted by the green (sensitive directions east-west) and  red (sensitive directions north-south) curves of  Fig.~\ref{fig:data}. The data spans the frequency range of $6\times 10^{-6}\units{sec^{-1}}\lesssim \omega \lesssim 5\times 10^{-3}\units{sec^{-1}}$. This data will be used to cast limits on ALPs with masses $ m_a\lsim 3 \times 10^{-18} \units{eV}$.

In addition to the data shown in Fig.~\ref{fig:data}, Ref.~\cite{BrownThesis} also provides us with a bound on the amplitude of a constant anomalous field. A constant anomalous field would be interpreted as a nearly massless ALP, $m_{a}<1.8\times 10^{-22}~\units{eV}$, so that ${\bf b}_n$ of Eq.~\eqref{eq:ALPsbfield} remains nearly constant throughout the measurement. This bound relies on the full 143 days of exposure, and is therefore stronger than the one cast from the 22 days of exposure of the longest uninterrupted measurement. Indeed, Ref.~\cite{Graham:2017ivz} has already cast a bound on neutron coupling to ultra-light ALPs from this result. However Ref.~\cite{Graham:2017ivz} has not accounted for the stochastic nature of the ALP field, which was recently shown to weaken bounds when taken to account~\cite{Centers:2019dyn} (see Sec.~\ref{sec:results} for a brief discussion, and Appendices~\ref{app:nonDeterministic} and~\ref{app:final} for a more complete one). We therefore recalculate that bound here, getting weaker results due to this effect. 

\end{enumerate}

\section{Analysis and Results}
\label{sec:results}

We now describe our analysis method for obtaining new constraints on the ALP parameter space using the comagnetometer measurements described above, and the results derived from the existing data. While there is no commonly accepted model for the ALP DM density and velocity distribution (see Ref.~\cite{Vergados:2016rlh} for several models), in order to cast bounds, one must choose a specific model. Here we have the average velocity of the ALPs relative to us, $|\left<{\bf v}\right>| = 232\units{km/sec}\sim 0.77\times 10^{-3}$ in natural units. For the density profile, we have assumed that the ALPs comprise of all DM in the galaxy have assumed an average ALP density of $\left<\rho_a\right>=\rho_{\rm DM}= 0.4 \units{\ GeV\ cm^{-3}}$.

Following Eq.~\eqref{CasperEQ}, the ALP-induced anomalous field is in the direction of the ALP velocity and linearly dependent on its size. The experimental sensitivity depends on the direction of the anomalous field, in relation to the detector's sensitive direction (encoded in $\hat{\bf \sigma}$ of Eq.~\eqref{CasperEQ}), which is different for the different datasets. The rotation of the earth rotates the  direction of sensitivity of the detector, creating ${\cal O}(1)$ daily modulation in the signal. The data of Ref.~\cite{BrownThesis} is the only one with a fine-enough resolution to measure this effect. The details of our daily modulation treatment are given in Appendix~\ref{app:directions}, and their application to the Ref.~\cite{BrownThesis} dataset is described in Appendix~\ref{app:final}.  

In the limit $m_a \rightarrow 0$, the signal can be constrained from the signal at the sidereal day frequency, $\omega=\Omega_{\rm SD}\simeq 2\pi/\units{day}$. ALPs in this limit can be thought of as a source to the original anomalous daily modulated field searched for in Refs.~\cite{KornackThesis,BrownThesis}. Indeed, the bound calculated by Ref.~\cite{Graham:2017ivz} assumes this. As we have mentioned, independently of the data in Fig.~\ref{fig:data}, Ref.~\cite{BrownThesis} presents their final values for a constant anomalous field, $|b_n^{\perp}|<3.7\times 10^{-33}\units{GeV}$ at $68\%$ C.L., corresponding at $95\%$ C.L. to $|b_n^{\perp}|<5.5\times 10^{-33}\units{GeV}$. 
Due to a long break in the middle of data-taking, the experiment was spanned over a period of $270\units{days}$, so that for masses of $m_a<2\pi/(270 \units{days})\simeq 1.8\times 10^{-22}\units{eV}$, the anomalous field would appear as constant throughout the experiment of Ref.~\cite{BrownThesis} (up to the effect of the daily modulation), and the result above can be used.

The naive plugging-in of $\rho_a=\left<\rho_{\rm DM}\right>,{\bf v}_a=\left<{\bf v}_a\right>,\theta_0=0$, and $E_a=m_a$, in Eq.~\eqref{eq:ALPsbfield}, to find the appropriate bound on the coupling neglects the stochastic nature of the ALPs, and is inaccurate by a factor of $\mathcal{O}(20)$. We will now discuss briefly the effects of the stochastic nature of the ALP field, though we leave the full discussion to Appendices~\ref{app:nonDeterministic} and~\ref{app:final}.

Non-relativistic ALPs  are  coherent over a period of 
\begin{equation}
\label{eq:coherence}
\tau_{a} \lesssim  \frac{2\pi}{m_a v_{\rm stochastic}^2} \simeq 8.9 \units{days}\left(\frac{10^{-14}\units{eV}}{m_a}\right)\,,
\end{equation}
Where we took $v_{\rm stochastic}=v_{\rm virial}=220~{\rm km/sec}$. For any measurement shorter than the coherence time, we can assume a single value was sampled for the velocity, the energy, the relative phase, and the density of the ALP field (the distributions can be found in Appendix~\ref{app:nonDeterministic}). For a measurement time $t_{\rm tot}=n\tau_a$, we assume $n$ random samples of the stochastic distributions should be summed upon. To get the bounds, we therefore run a simple Monte Carlo (MC) simulation in which we sample the distributions appropriately and derive a $95\%$~C.L. bound. 

The stochastic nature of the ALPs allows for the possibility that we are in a region where the ALP field is uncharacteristically small. However, it is enough to draw few samples from the distributions to decrease this effect. This implies an $\mathcal{O}(3)$ improvement of the sensitivity when transitioning from $t_{\rm tot}=\tau_a$ to $t_{\rm tot}\sim 5\tau_a$. As explained in Appendix~\ref{app:nonDeterministic}, a more detailed analysis could be made with more complete-data, achieving a further improvement to the bound which scales as $t_{\rm tot}^{1/4}$ for the Signal-to-Noise-Ratio (SNR) at periods longer than the coherence time; however, with the current data available, that additional improvement could not be achieved.

The 7 runs of Ref.~\cite{VasilakisThesis} were spread over a period of $\sim 100$~days, and the longest of them took about $\sim 8$ days. Therefore, for masses $10^{-14}\units{eV}\lsim m_a\lsim 10^{-15}\units{eV}$, the signal would be incoherent for the 100 days, despite being coherent for the entirety of any specific run. Another non-trivial detail is the calibration procedures that were done during the measurements -- which took about ${\cal O}(50\%)$ of the total measurement time (during which no data was recorded). In general, all three datasets shown in Fig.~\ref{fig:data} have gone through several processing procedures, which we do not know in full details. These complications give rise to some uncertainties, which we discuss in further details in Appendix~\ref{app:final}.

\begin{figure*}[th!]	
	\includegraphics[width=0.9\textwidth]{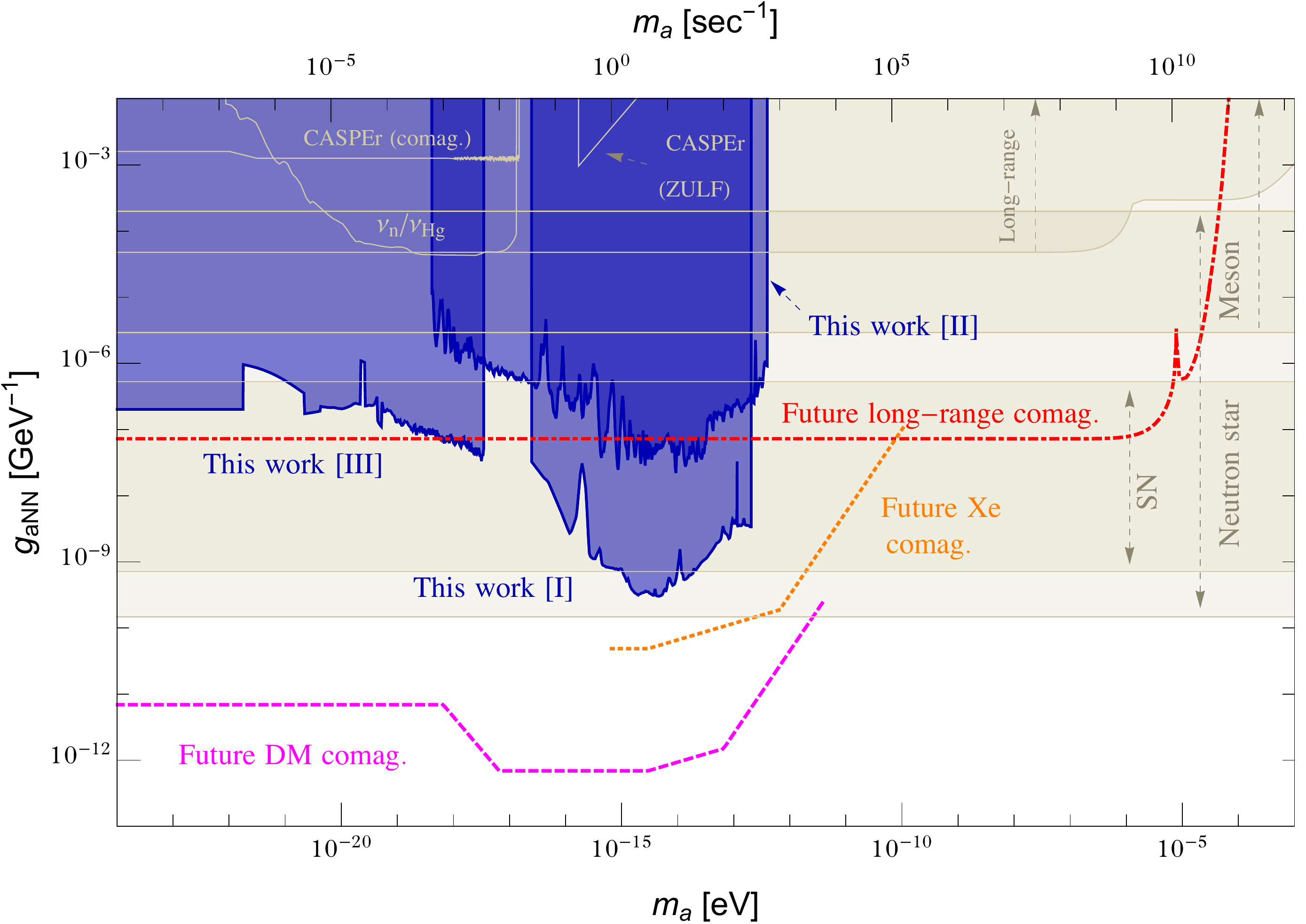}
	\caption{\label{fig:gannbounds}
	Constraints and projected reach for ALP-neutron couplings. The shaded blue regions represents the $95\%$ C.L. bounds derived in this paper from datasets I~\cite{VasilakisThesis}, II~\cite{KornackThesis} and III~\cite{BrownThesis};  the bound derived from dataset III continues to arbitrarily small ALP masses. The sudden increase of the bound at ultra-light masses is due to longer measurement-time being available at those masses, and not an increased sensitivity at low frequencies.
	The shaded `long-range' region comes from the non-observation of deviations from the gravitational $1/r^2$ at short distances~\cite{Adelberger2007},  together with the bound from long-range spin-dependent interactions~\cite{VasilakisThesis}.  The `$\nu_n/\nu_{\rm Hg}$' shaded region comes from Ref.~\cite{Abel:2017rtm}, which compared the effect of anomalous DM axion fields on Hg and neutrons. Similarly, the `CASPEr (comag.)' region is excluded by the non-observation of the effect of anomalous DM axion fields on ${}^1$H and ${}^{13}$C~\cite{Wu:2019exd}. The `CASPEr (ZULF)' shaded region indicates the phase-I bound of that experiment~\cite{Garcon:2019inh} which looks for anomalous fields by utilizing NMR methods. The last three bounds were recently corrected by Ref.~\cite{Centers:2019dyn} which accounts for the previously ignored stochastic nature of the ALP field, and we use their corrected results in this figure. The `neutron star' band indicates the constraints from neutron star cooling considerations~\cite{Beznogov:2018fda}. The `SN' band depicts cooling bounds from supernova SN1987a~\cite{Carenza:2019pxu}. The `meson' band is the model-dependent bound from searching for invisible meson decays~\cite{Essig:2010gu}. The dashed magenta, dotted orange and dot-dashed red curves  indicate future reach of our proposed improved experimental setups; for further details, see main text.
}
\end{figure*}

\begin{figure*}
	\includegraphics[width=0.9\textwidth]{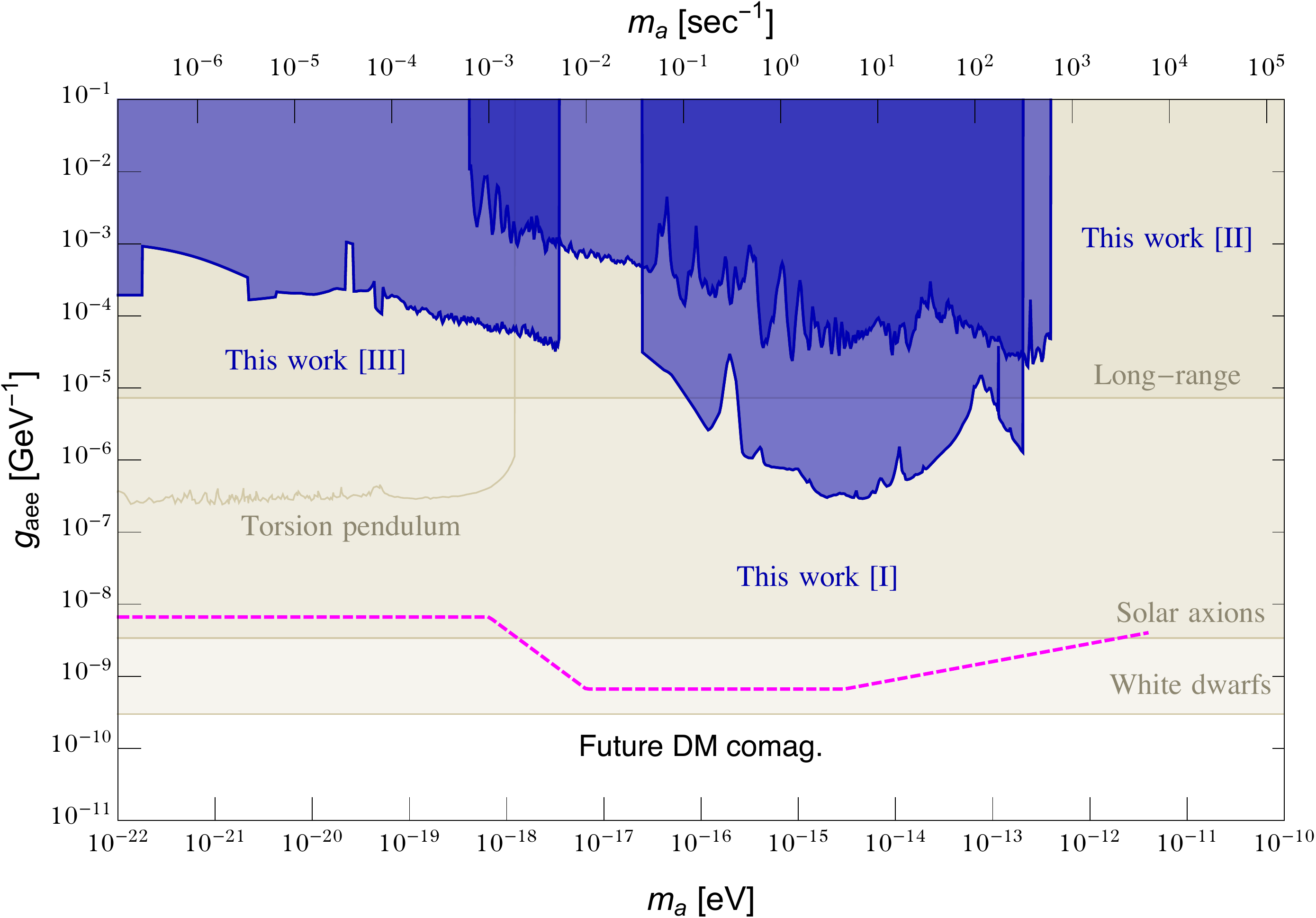}
	\caption{\label{fig:gaeebounds}
Constraints and projected reach for ALP-electron couplings. The shaded blue regions represent the $95\%$ C.L. bounds derived in this paper from datasets I~\cite{VasilakisThesis}, II~\cite{KornackThesis} and III~\cite{BrownThesis}, the third of which continue to arbitrarily small ALP masses. The sudden increase of the bound at ultra-light masses is due to longer measurement-time being available at those masses, and not an increased sensitivity at low frequencies. The shaded `long-range' region represents the constraint from searching for new long range interactions~\cite{Terrano:2015sna}. The `Torsion-Pendulum' region represents the bound from the search for the anomalous field sourced by ALPs, interacting with polarized electrons of a so called ''spin-pendulum"~\cite{Terrano:2019clh}. The shaded `solar axions' region is excluded by the solar axion search of the LUX collaboration~\cite{Akerib:2017uem}. The shaded `white dwarfs' region is excluded by considering the effects axions would have on white dwarfs as a new cooling mechanism~\cite{Bertolami:2014wua}. The magenta dashed curve describes the future reach of an improved comagnetometer setup we propose; see main text for further details. 
}
		\end{figure*}

In general, more of the technical details for the procedure that we use to derive our constraints are described in the Appendices, with the final procedure found in Appendix~\ref{app:final}. Our results for the constraints on the ALP-neutron and ALP-electron couplings  are shown as blue shaded regions in Figs.~\ref{fig:gannbounds} and~\ref{fig:gaeebounds},  respectively.

In Fig.~\ref{fig:gannbounds}, the region labeled as `long-range' represents the merging of two separate bounds from the non-observation of new long range interactions~\cite{VasilakisThesis,Adelberger2007}. The `$\nu_n/\nu_{\rm Hg}$' region is excluded from not measuring anomalous fields in a system of mercury atoms and free neutrons~\cite{Abel:2017rtm}, and the `CASPEr (ZULF)' region is excluded by the phase~I run of this low-frequency NMR  experiment~\cite{Garcon:2019inh}. The bound from the CASPEr ZULF comagnetometer experiment is presented as the `CASPEr (comag.)' region~\cite{Wu:2019exd}\footnote{We note that the CASPEr ZULF comagnetometer is conceptually similar but different to the noble-alkali comagnetometer discussed in this paper. Much like the experiment that produced the $\nu_n/\nu_{\rm Hg}$ line presented in Fig.~\ref{fig:gannbounds}, it measures the frequency induced by anomalous fields in two different spins, ${}^{13}$C and ${}^1$H in this case -- however they are not strongly coupled as in the case of the noble-alkali comagnetometer presented in this paper.}. The last three exclusion regions were recently corrected by Ref.~\cite{Centers:2019dyn} which accounts for the previously ignored stochastic nature of the ALP field, and we use their corrected results in this figure. The `neutron star' and `SN' shaded regions indicate the stellar constraints from neutron star cooling~\cite{Beznogov:2018fda}  and supernova SN1987a~\cite{Carenza:2019pxu} (a recent recalculation of Ref.~\cite{Chang:2018rso}). The `meson' shaded region is a model-dependent constraint, arising from the new decay channels that axions would introduce in meson decays~\cite{Essig:2010gu}.  The dotted orange, the dashed magenta, and the dot-dashed red curves show our future projections, as explained in the next section. 

In Fig.~\ref{fig:gaeebounds}, the `white dwarfs' and `solar axions' shaded regions indicate astrophysical constraints coming from the new cooling mechanism axions would introduce in white dwarfs~\cite{Bertolami:2014wua}, and the non-observation of solar axions by the LUX experiment~\cite{Akerib:2017uem}, respectively. The `long-range' region presents the bound from looking for long-range spin-dependent interactions~\cite{Terrano:2015sna}. The `torsion-pendulum' region presents the bound from the search for the anomalous field sourced by ALPs, interacting with polarized electrons of a so called ``spin-pendulum''~\cite{Terrano:2019clh}. The magenta dashed curve shows our future projections of a dedicated comagnetometer experiment, as explained in the next section. 

As is evident, for ALP-neutron couplings, our new derived bounds from old data provide the strongest terrestrial constraints to date over a broad range of masses, providing a complementary probe to stellar constraints. Further improvements and deep reach into uncharted parameter space should be made possible with future experimental improvements, as we now detail.

\section{Future Improved Experiments}
\label{sec:future}

The concept of the alkali-noble comagnetometers exists for over a decade and shows great promise, however relatively little work on the topic discussed here has been performed. We now outline several possible directions for future improvement, which could enhance the sensitivity of these systems to relic ALPs. We describe three realistic experimental setups for improved sensitivity. The Hot Vapors Laboratory of the Quantum Optics Center in Israel is currently building two comagnetometers that will implement some of the ideas presented below. Our projected sensitivity curves for these realistic future experimental setups are depicted by the dashed curves in Fig.~\ref{fig:gannbounds} and Fig.~\ref{fig:gaeebounds}.

\subsection{Dedicated DM Search}

The simplest way to improve the bounds extracted in this paper is to improve the detector and by performing specific analysis  for DM.   The expected reach is shown by the dashed magenta curves in Fig.~\ref{fig:gannbounds} and  Fig.~\ref{fig:gaeebounds}. The predicted constraint realistically assumes a 30-day dedicated run with an $\mathcal{O}(5-10)$ improvement in the signal-to-noise ratio (SNR) of the detector compared to Ref.~\cite{VasilakisThesis}, {\it i.e.} assuming $1.4\units{pico{\rm G}/\sqrt{Hz}}$ noise spectrum density. We further assume the noise to increase by an order of magnitude at the lower frequencies. We have assumed a moderate increase in the polarization of the helium atoms, leading to a compensation frequency ($\omega_c\equiv \gamma_{\rm He}B_c$) of $100\units{sec^{-1}}$, as well as a small increase in the alkali decay rate ($R_e/q\gsim\omega_c=100\units{sec^{-1}}$). As discussed in Appendix~\ref{app:dynamics}, when the frequency of the anomalous fields, $\omega$, increases, the SNR of the detector decreases as a function of  $\omega / (R_e/q)$ and $\omega/\omega_c$. In this neutron-ALP interaction projection, we have therefore taken the SNR of the detector to linearly decrease after reaching $100\units{sec^{-1}}$. For the electron-ALP interactions, the loss of SNR is less steep, and we have therefore neglected this effect. All of these improvements rely on advanced techniques which are currently being tested.

One of the detectors being built has two probe beams, and it is planned to implement a control measurement ({\it e.g.} by exiting the compensation point for short intervals during which sensitivity to  noise from magnetic fields is present). We believe that our background subtraction can introduce significant additional improvements compared to what is currently shown in Fig.~\ref{fig:gannbounds} and Fig.~\ref{fig:gaeebounds}. We also expect to be able to have a marginal improvement at times longer than $\tau_{a}$, that scales as $\sqrt[4]{t_{\rm tot}}$, as discussed in Appendix~\ref{app:nonDeterministic}. Ref.~\cite{VasilakisThesis}  has already shown that even a partial control measurement can introduce an $\mathcal{O}(10)$ improvement for some frequencies, and it is therefore likely that the final reach of such an experiment could be even greater than that presented here.
 
An additional improvement is expected through complete 3D knowledge of directionality of the measured anomalous field. Since the directional properties of the experimental noise are currently unknown, we do not include potential improvement from a multi-directional search in our projected reach. Techniques to measure the entire 3D vector of the anomalous field are, however, currently being studied and will enable the complete knowledge of the ALP field directionality.  
If a sharp peak is found at some frequency, directional detection schemes in different laboratories would allow testing whether the measured signal is sourced out of earth, which will inform us in the question of its DM origin. 
 
\subsection{Change of Atoms}

While the ${}^3$He-K comagnetometer can achieve strong bounds on the interactions between ALPs and neutrons, it cannot probe ALP-proton interactions due to the absence of a proton component in the ${}^3$He spin (see Appendix~\ref{app:nuc} for further details). Changing the identity of the atoms  the comagnetometer can not only affect the sensitivity but also further enable the probing of the ALP-proton coupling, which can be much larger than the ALP-neutron coupling~\cite{diCortona:2015ldu}.  

Several options for variations in the atoms exist. One, currently under study,  is the use of ${}^{21}$Ne as an alternative to  ${}^3$He.  A second, more readily available is the use of a Xenon isotope, ${}^{129}$Xe or ${}^{131}$Xe, paired with Rb alkali atoms.
The Xe-Rb interactions trigger a large relaxation rate for the rubidium.  As a consequence, in order to reach reasonable polarizations, a cell with xenon isotopes must have a significantly lower pressure  compared to a cell with ${}^3$He atoms. Since the noise cancellation is also sensitive to the density of the noble gas, $\delta B/B_c\propto 1/n_{\rm noble}$, this would naively impede the cancellation. However, the interaction of the Rb and Xe is about $\mathcal{O}(100)$ stronger than that of K and ${}^3$He, and additionally the pumping of the Xe isotopes can reach $\mathcal{O}(10\%)$ polarization compared to the $\mathcal{O}(2\%)$ polarization of the ${}^3$He atoms. Thus an order of magnitude increase in the compensation frequency can be expected.

The decay rate of the electronic spin is increased by orders of magnitude compared to the ${}^3$He-K comagnetometer, which would naively suppress the signal as can be seen from Eq.~\eqref{eq:sigperp}. However, the leading order contribution from magnetic noise is suppressed by the same factor, so the SNR due to magnetic noises is unaffected.\footnote{The leading contribution for magnetic noise from oscillating magnetic fields is presented in Appendix~\ref{app:dynamics}, Eq.~\eqref{eq:highfK}. However, the magnetic noises stemming from inaccurately tuning the system to the compensation point also scale as $1/R_e$. The derivation of these corrections are beyond the scope of this paper, and their description can be found in Refs.~\cite{KornackThesis,VasilakisThesis,BrownThesis}.} Depending on the precise origin of the detector noise---whether sourced by SM magnetic fields or something else, such as noise in the lasers---this suppression of both signal and noise may or may not fully cancel. For this reason, we use a conservative projection assuming that the detector will experience a constant spectral density noise of  $0.1\units{n{\rm G}/\sqrt{\rm Hz}}$. As in the dedicated DM search case, the increase in compensation field and decay rate has been translated to the dynamical suppression factors discussed in Appendix~\ref{app:dynamics} appearing at higher frequencies compared to the existing experiments analyzed here. For the assumed $\omega_c=10^3\units{sec^{-1}}$ this translates to a linear decrease in the sensitivity starting at $m_a=10^3\units{sec^{-1}}$. 

Our projection is shown by the orange dotted curve in Fig.~\ref{fig:gannbounds}. The noise of the detector at low frequencies is extremely hard to predict and thus the projected bounds are given for masses $m_a>\units{sec^{-1}}$. Once again we assume a 30~day  run period. The reach of the Xe-Rb comagnetometer described here can also be cast for ALP-proton couplings, with similar sensitivity to the ALP-neutron ones, thus providing a complementary probe. The sensitivity of this detector to ALP-electron interactions are not-competitive with existing bounds and are therefore  not shown.
Finally, we comment that the Xenon detector may be further improved by the simultaneous use of two Xenon isotopes, as will be demonstrated in future work. 
 
\subsection{Long Range Spin-Dependent Interaction Search}

In the experiment of Ref.~\cite{VasilakisThesis}---whose data was analyzed in this work---a helium-potassium comagnetometer is used to measure long-range spin-dependent interactions via an independent sample of highly polarized ${}^3$He gas, placed in proximity to the detector. This is equivalent to searching for the effective anomalous field generated by the highly polarized source of ${}^3$He. In this case, however, the interaction is only modulated if the directionality of the source sample is modulated, giving control over the frequency of the sought signal. A future improved long-range spin-dependent interaction search would enable significant reach into ALP parameter space. We note that this experiment probes the existence of ALP interactions regardless to whether they are a component of DM or not.

The $\mathcal{O}(10)$ improvement in the SNR of the planned future ${}^3$He-K comagnetometer would give an $\mathcal{O}(3)$ improvement on the bounds. Changing the geometry can further enhance the reach.  A factor of ${\cal O}(10)$ improvement on the bound can be achieved by placing the sample source at a distance of 10~cm from the center of the comagnetometer, rather than the 50~cm distance of Ref.~\cite{VasilakisThesis}. The specialized detector currently being built is much smaller than that of Ref.~\cite{VasilakisThesis} and should thus allow for this close placement of the source sample. Such a setup would then allow to probe masses a factor of 5 heavier than those probed in Ref.~\cite{VasilakisThesis}.

A final further planned improvement, which should yield an additional enhancement of reach by a factor of ${\cal O}(20)$, is the use of xenon in one of its non gaseous phases (xenon ice, liquid xenon, or xenon snow) as the source material. By taking a cell of 5~cm radius filled with non-gaseous xenon, with an achievable polarization of $50\%$, a substantially increased amount of polarized spins can be obtained. We note that the typical constructed cells of such polarized material are usually much smaller than this. However, Ref.~\cite{Garcon:2017ixh} has discussed the use of such cells in one of their future phases, and cells of size $300 \units{mL}$ (oddly shaped, but close in volume to a $5\units{cm}$ radius spherical cell), have already been used in Ref.~\cite{FreemanThesis} with 34\% polarization of Xe (though since the thawed gas is said to lose about $50-80\%$ of its initial polarization, we expect substantial improvement is possible in the absence of thawing). The greatest challenge of using the existing cell technology is that these cells are commonly housed in strong magnets which could ruin the comagnetometer's shields, making the placing of the comagnetometer $10\units{cm}$ away from the anomalous spin source a challenging task. Preliminary investigation however implies that these issues may be solved in the future, and our projected reach for this future experiment are shown by the dot-dashed red curve in Fig.~\ref{fig:gannbounds}. We note that this type of experiment has no independent sensitivity to $g_{aee}$.

\section{Summary}

Comagnetometers present an innovative and under-utilized avenue to probe ultra-light ALPs. With current setups far from optimization, and sensitivity spanning many decades of ALP masses, down to  fuzzy dark matter~\cite{Hu:2000ke,Lee:2017qve,Hlozek:2014lca} 
 masses of $\mathcal{O}(10^{-22} \units{eV})$, comagnetometers hold great promise to detect relic ALPs. 
In this paper we have presented the foundation to enable current and  future searches using comagnetometers to constrain and detect such ultra-light ALPs. Using publicly available partial comagnetometer data, we are able to place meaningful constraints on ALP couplings to neutrons and electrons, including, in the case of ALP-neutron interactions, the strongest terrestrial constraints to date over a broad range of masses, demonstrating the power of our approach. With future improvements to the experimental setup---the implementation of which is already underway---many different and interesting searches can be performed, with prospects to cut deep into unchartered ALP parameter space in the near future.

\mysections{Acknowledgments}
We thank Or Katz, Roy Shaham, Ran Fischer, Tal David and Shmuel Nussinov for helpful discussions. We also thank Michael Romalis for helpful information on the data, and Gary Centers for his help with the formulation of the treatment of the stochastic nature of the ALP field.
IB is grateful for the support of the Alexander Zaks Scholarship, The Buchmann Scholarship and of the Azrieli Foundation.
 The work of YH is supported by the Israel Science Foundation (grant No. 1112/17), by the Binational Science Foundation (grant No. 2016155), by the I-CORE Program of the Planning Budgeting Committee (grant No. 1937/12), by the German Israel Foundation (grant No. I-2487-303.7/2017), and  by the Azrieli Foundation.  EK is supported by the Israel Science Foundation
(grant No. 1111/17), and by the I-CORE Program
of the Planning Budgeting Committee (grant No.
1937/12). EK and TV are supported by the Binational Science Foundation
(grant No. 2016153).  TV is further supported in part by the Israel Science Foundation-NSFC
(grant No. 2522/17), by the European Research Council (ERC) under the EU Horizon 2020 Programme (ERC- CoG-2015 - Proposal n. 682676 LDMThExp), and by a grant from the Ambrose
Monell Foundation, given by the Institute for Advanced Study.

 \appendix
\section{Detailed Derivation of the Comagnetometer's Steady State Behavior}\label{app:comag}

\begin{table}[]
\begin{tabular}{|c||c|c|c|}
\hline Scale	& Ref.~\cite{VasilakisThesis} & Ref.~\cite{KornackThesis}	& Ref.~\cite{BrownThesis}	\\ \hline\hline
$\Gamma_{\rm K}\units{[sec^{-1}]}$	& $63$	& $79$	& $68$\\ \hline
$\omega_{\rm K}\units{[sec^{-1}]}$	& $-24$	& $-19$	& $-48$	\\ \hline
$\omega_{\rm He}\units{[sec^{-1}]}$	& $-110$	& $-32$	& $-53$	\\ \hline
$\omega_{\rm K-He}\units{[10^3\ sec^{-1}]}$	& $260$ 	& $330$	& $230$		\\ \hline
$\omega_{\rm He-K}\units{[10^{-3}\ sec^{-1}]}$	& $9.8$	& $1.9$	& $11$		\\ \hline
$\omega_{\rm fast}\units{[sec^{-1}]}$	& $-4.2$	& $-13$	& $-13$		\\ \hline
$\omega_{\rm slow}\units{[sec^{-1}]}$	& $-130$	& $-38$	& $-88$		\\ \hline
$\Gamma_{\rm fast}\units{[sec^{-1}]}$	& $53$	& $72$	& $36$		\\ \hline

$\Gamma_{\rm slow}\units{[sec^{-1}]}$	& $10$	& $7$	& $32$		\\ \hline
\end{tabular}
\caption{The important scales which appear in Eq.~\eqref{eq:matrix}, calculated using the values of Table~\ref{table:raw}. The $-\Gamma_{\rm fast}+i\omega_{\rm fast},-\Gamma_{\rm slow}+i\omega_{\rm slow}$ are the eigenvalues of the matrix in Eq.~\eqref{eq:matrix}. The ``fast'' (``slow'') subscript corresponds to the mode that far from the compensation point was the alkali (noble) spins' mode.}
\label{table:times}
\end{table}

This appendix delves into the detailed description of the comagnetometer's steady state equations, leaving the time-dependence of the system to Appendix~\ref{app:dynamics}. Our goal is to present some of the details of the derivation of Eqs.~\eqref{eq:Sdot} and~\eqref{eq:Kdot}, and then discuss the derivation of Eq.~\eqref{eq:sigperp}.

The spin of each individual potassium atom is composed of an electronic spin-$1/2$  and a nuclear spin-$3/2$ configurations. As a consequence, the  Bloch equations which describe the spin degrees of freedom as 3-vectors, cannot be  naively used, and a more complex density matrix formalism seems necessary. To simplify the situation, it is possible to integrate over the nuclear degrees of freedom to reach an effective spin-$1/2$ system which then allows one to use the Bloch equations with some of the constants modified to account for the integrated-out degrees of freedom. This is precisely the method used to arrive at Eq.~\eqref{eq:Sdot} (and similarly, Eq.~\eqref{eq:anomalous}), with $q$, the slowing down factor encapsulating the nuclear degrees of freedom.  Generally, $R_e$ is not isotropic,  and there is a much faster decay rate in the directions perpendicular to the magnetic field, however, in the so called SERF regime which we are working in, this anisotropy can be neglected~\cite{Allred2002}. Finally, the ${}^3$He are spin-$1/2$ atoms with their spin stemming entirely from the neutron in the nucleus~\cite{Stadnik:2014xja}, and thus the Bloch equations are immediately applicable for them.

Let us consider approximate solutions to Eqs.~\eqref{eq:Sdot},\eqref{eq:Kdot}.   Six degrees of freedom are at play: 3 from ${\bf S}_{\rm K}$, and 3 from ${\bf S}_{\rm He}$. In standard operating procedure, all of the magnetic fields (external, as well as those induced by the atoms on each other) are approximately aligned with the $\hat{\bf z}$ direction, (which is the pump beam direction as well), so there are no transverse polarizations. As a consequence, at leading order there are only 2 degrees of freedom, corresponding to the $\hat{\bf z}$ polarizations. Moreover, after time $t\sim(3/R_{\rm pu}^{\rm eff})$, the system reaches a steady state, so that at leading order in the misalignments,
\begin{equation}
S^{z(0)}_{\rm K}=\frac{R_{\rm pu}}{2R_e},
\end{equation}
\begin{equation}
S^{z(0)}_{\rm He}=\frac{R_{\rm pu}^{\rm eff}}{R_{\rm He}+R_{\rm pu}^{\rm eff}}\cdot\frac{R_{\rm pu}}{2R_{e}}.
\end{equation}

As can be seen in Eqs.~\eqref{eq:Sdot},\eqref{eq:Kdot}, the next order corrections would only contribute to the transverse components, as the  leading effect of a misalignment is to rotate the spins without changing their absolute value, {\it i.e.} we expect $S^{\perp(1)}\sim S^{z(0)}\cdot \sin(\theta)$, with $\theta$ representing the misalignment, while the longitudinal component receives no correction to order ${\cal O}(\theta)$. 

We may thus conclude that the first order equations have four real degrees of freedom corresponding to the four transverse components. These equations can be written more compactly by complexifying a general 3-vector ${\bf v}=(v^x,v^y,v^z)$, writing it as $v^{\mathbb{C}}=v^x+iv^y$, and $v^z$ instead. The first order equations for the transverse components can therefore be written as a $2\times 2$ linear ordinary differential equation with constant coefficients and inhomogeneous terms,
\begin{equation}
\label{eq:matrix}
\begin{aligned}
&\left(\begin{array}{c} \dot{S}^{\mathbb{C}}_{\rm K} \\\dot{S}^{\mathbb{C}}_{\rm He}\end{array}\right)=\left(\begin{array}{cc}i\omega_{\rm K}-\Gamma_{\rm K} &- i\omega_{\rm K-He } \\- i\omega_{{\rm He-K}} & i\omega_{\rm He}\end{array}\right)\left(\begin{array}{c}S^{\mathbb{C}}_{\rm K} \\S^{\mathbb{C}}_{\rm He}\end{array}\right)
\\
&-\left(\begin{array}{c}\frac{i\gamma_e S^{z(0)}_{\rm K}}{q}B^{\mathbb{C}} \\ i\gamma_{\text{He}} S^{z(0)}_{\rm He}B^{\mathbb{C}}\end{array}\right)-\left(\begin{array}{c}iS^{z(0)}_{\rm K} b^{\mathbb{C}}_e/q \\i\gamma_{\rm He}S^{z(0)}_{\rm He}b_n^{\mathbb{C}}/\gamma_n\end{array}\right)\,.
\end{aligned}
\end{equation}
Here $\omega_{\rm K}=\gamma_e B^z/q+2\gamma_e\lambda \mu_{\rm He} S^{z(0)}_{\rm He}/q$  and $\omega_{\rm He}=\gamma_{\text{He}}B^z+2\gamma_{\text{He}}\lambda \mu_{\rm K} S^{z(0)}_{\rm K}$ are the precession frequency of the transverse components around the $\hat{\bf z}$ direction for the potassium  and helium atoms' spins respetively. $\Gamma_{\rm K}=R_e/q$ is the time scale typical for a precession of the potassium atoms' spins to decay. The off diagonal term $\omega_{\rm K-He }=2\gamma_e \lambda\mu_{\rm He} S^{z(0)}_{\rm K}/q$ ($\omega_{{\rm He}-K}=2\gamma_{\text{He}} \lambda\mu_{\rm K} S^{z(0)}_{\rm He}$) represents the rotation of the zeroth order $\hat{\bf z}$ potassium (helium) polarization around the magnetic field generated by the transverse helium (potassium) polarization. The inhomogeneous terms come from the rotation of the zeroth order $\hat{\bf z}$ polarizations around the small transverse magnetic and anomalous fields. The typical values for the scales presented in this equation can be found in Table~\ref{table:times}. In the above, all terms proportional to the timescales $R_{\rm He},R_{\rm pu}^{\rm eff}$ were neglected as they are much slower than any other relevant rate. Additionally, anomalous fields in the $\hat{\bf z}$ direction were neglected as they are significantly smaller than the external magnetic fields at play along this direction.

The goal of the detector is to measure the transverse anomalous fields, and Eq.~\eqref{eq:matrix} implies that the transverse magnetic fields have a similar effect  on the polarizations and therefore act as background. However, because the terms are not exactly the same, and the detector only measures the potassium's spin, by tuning the magnetic fields along the  $\hat{\bf z}$ direction, it is possible to greatly decrease that background.
To see this, let us consider the limit $b_n^{\mathbb{C}}=b_e^{\mathbb{C}}=0$. The steady state solution, $\dot{S}_{\rm K}^{\mathbb{C}}$=$\dot{S}_{\rm He}^{\mathbb{C}}=0$,  then implies that independently of the size and direction of the transverse magnetic fields, if
\begin{equation}
\label{eq:compensationpoint}
B^z=B_c\equiv -2\lambda \mu_{\rm K} S^{z(0)}_{\rm K}-2\lambda \mu_{\rm He} S^{z(0)}_{\rm He},
\end{equation}
then $S^{\mathbb{C}}_{\rm K}=0$. In other words, if the external magnetic field's $\hat{\bf z}$ component is tuned to $B_c$, then transverse magnetic fields have no first order effect on the steady-state transverse potassium polarization. When the system is in the state where $B^z=B_c$, it is said to be in the {\it compensation point}. 

We note here that away from the compensation point, the sensitivity to non-anomalous transverse magnetic fields is restored.
For this reason, magnetic shielding is crucial, allowing for the stabilization of the system around that point.  We also note that as explained in section 3 of Ref.~\cite{BrownThesis}, the $\mu$-metal shields used in such systems do not shield anomalous fields.\footnote{In short, $\mu$-metal magnetic shields do not respond to ${\bf b}_n$, while their response to ${\bf b}_e$ generates an oppositely directed magnetic field, which a comagnetometer tuned to the compensation point would be insensitive to.} 

At the compensation point, the sensitivity to constant anomalous fields is easily found by taking the steady state condition again, $\dot{S}^{\mathbb{C}}_{\rm K}=\dot{S}^{\mathbb{C}}_{\rm He}=0$, and solving for $S^{\mathbb{C}}_{\rm K}$ and $S^{\mathbb{C}}_{\rm He}$, one finds 
\begin{eqnarray}
\label{eq:romsigCompK}
S_K^{\mathbb{C}} &=& \frac{iS^{z(0)}_{\rm K}}{R_e}\left(\frac{\gamma_e}{\gamma_n}b_n^{\mathbb{C}}-b_e^{\mathbb{C}} \right),
\\
\label{eq:romsigCompHe}
S_{\rm He}^{\mathbb{C}}&=&-\frac{\gamma_n B^{\mathbb{C}}+b_n^{\mathbb{C}}}{2\gamma_n \lambda  \mu_{\rm He}}-\frac{\mu_{\rm K}}{\mu_{\rm He}}\frac{iS_{\rm K}^{z(0)}}{R_e}\left(\frac{\gamma_e}{\gamma_n}b_n^{\mathbb{C}}-b_e^{\mathbb{C}}\right)\,,
\end{eqnarray}
where we only took the leading order in $1/R_e$ which is the fastest rate in this setup. Note that Eq.~\eqref{eq:romsigCompK} is equivalent to Eq.~\eqref{eq:sigperp} from the main text, up to notation.  From the above one sees that indeed the alkali's transverse magnetization is insensitive to the external magnetic field, while that of the helium-3 is (thereby allowing the cancelation in the alkali system).

\section{The Dynamical Response of the Comagnetometer}\label{app:dynamics}

Our goal is to understand the dynamical response of a system described by Eq.~\eqref{eq:matrix}.
When an anomalous field is rapidly oscillating, the spins in the system are unable to follow the changes sufficiently fast, and therefore the signal is suppressed. Additionally, at the compensation point, the nuclear spin must  follow the outside magnetic field in order cancel the total magnetic field felt by the alkali. This does not occur for a rapidly varying field and as a result, the alkali spins will be affected by the external magnetic fields, implying a  subpar  noise cancellation. It is thus clear that the dynamical response to changing fields is crucial, and in this appendix we explain how the effects of abrupt changes and oscillating fields on the comagnetometer can be calculated, summarizing the main results of the calculation.

The solution to a linear non-homogeneous $2 \times 2$ ODE such as Eq.~\eqref{eq:matrix} is composed of  homogeneous and  inhomogeneous contributions.  In the steady-state limit and after a sufficiently long time (compared with the inverse decay rate of the system to be discussed below), the homogeneous solution is exponentially small and the system is described by the inhomogeneous contribution, which in our case is controlled by the (possibly oscillating) fields.  We stress that near the compensation point, and for low magnetic frequencies, this part of the alkali's solution is {\it insensitive} to the non-anomalous fields (see Eq.~\eqref{eq:romsigCompK};  for higher frequencies to be discussed below, this is no longer true, however our treatment here of abrupt changes in the fields remains intact).
Conversely, before reaching the steady-state regime, the homogeneous solutions, determined by initial conditions, play an important role.  Time dependence in the system therefore enters in two distinct manners:  (i) abrupt changes drive the system away from the steady-state solution and can be described via initial conditions which alter the homogeneous solutions and (ii) oscillatory fields, the response to which is described within the steady-state regime, and shows up in the inhomogeneous part of the solution.
We now discuss each of these contributions separately. 

\subsection{The Homogeneous Solution: Response to Abrupt Changes}

Relevant abrupt changes in the comagnetometer system would  appear as sudden variations in the non-anomalous {\it transverse} magnetic fields, which show up in  the first inhomogeneous  terms of Eq.~\eqref{eq:matrix}.\footnote{Abrupt changes can also appear in $B_z$, however these alter the solution only at next order in perturbation theory, with corrections of order $\delta B_z\cdot B_{\perp}/B_{\rm comp}^2$.}  
Such changes keep the compensation point intact [see Eq.~\eqref{eq:compensationpoint}], however at short time scales, the helium-3 is too slow to align with the new magnetic fields and hence its influence on the alkali (through an induced magnetic field) does not cancel external magnetic field.   During this time, the system is susceptible to these fields and the sensitivity to anomalous fields is impaired.  

How is the above picture reflected in the solutions to Eqs.~\eqref{eq:Sdot} and~\eqref{eq:Kdot} and subsequently Eq.~\eqref{eq:matrix}?
While the numerical solutions which corresponds to the above discussion is easy to derive, the analytic solution is rather cumbersome and non-informative and hence we do not reproduce it here. 
Instead let us explain the important effects in the solution.

As discussed above, sufficiently close to the compensation point, the inhomogeneous part of the alkali's solution (which essentially describes the late-time steady-state behavior of the system) is largely independent of the non-anomalous fields, and therefore  sudden changes (typically relevant in low-frequency magnetic modes) in those  fields can only appear in the homogeneous contribution.   This is not the case for the helium-3, the inhomogeneous solution of which depends on all magnetic fields [Eq.~\eqref{eq:romsigCompHe}] and therefore alter upon a sudden change in the external fields.   Meanwhile, the homogeneous part of the solution (of both atoms) depends only on the parameters of the system but not the external fields, however, their coefficients (describing the most general solution), which are determined via the initial conditions, may regain such dependence. Since the two magnetometers are coupled (as is apparent through the non-diagonal terms in Eq~\eqref{eq:matrix}), the homogeneous solution of the two atoms is not aligned with the alkali and helium-3 modes.  The dependence of the helium-3 solution on the non-anomalous fields therefore influence the coefficients and  remains important so long as the homogeneous solution is not exponentially diluted (i.e. before the system reaches steady-state).  From that point of view, the homogeneous part of the solution entails the system's ability to respond to the sudden changes in the inhomogeneous terms.

In the  discussion so far, the system was described at short timescales, before it can reach its steady-state behavior.  Let us now estimate this timescale.  If not for the coupling of the two spin ensembles, there would be two distinct modes, one for the alkali and one for the noble gas.  The rate with which the noble gas's mode decays in such a case is longer than that of the alkali by many orders of magnitude. The interaction between the atoms mix the two spin modes and the resulting system is described by two new eigen-modes with two new respective eigenvalues. Since we mostly care about how long it takes the system to reach equilibrium, it is sufficient to discuss the slower decay rate, $\Gamma_{\rm slow}$.  Neglecting the rates, $R_{\rm K-He}, R_{\rm He-K}$ ( which would have been the real components of the off-diagonal terms of Eq.~\eqref{eq:matrix}), and $R_{\text{He}}$ (which are mostly irrelevant in the systems at hand), one finds for $\delta\omega\equiv \omega_{\rm K}-\omega_{\rm He} \ll \Gamma_K, \sqrt{\omega_{\rm He-K}\omega_{\rm K-He}}$,
\begin{equation}
\Gamma_{\rm slow}\simeq\frac{\Gamma_{\rm K}}{\delta\omega}\cdot\frac{\omega_{\rm He-K}\omega_{\rm K-He}}{\delta\omega}\,.
\end{equation}
 The above is only an order of magnitude smaller than the (mostly) alkali mode's decay rate in typical systems for which  $\Gamma_{\rm K}\simeq\sqrt{\omega_{\rm He-K}\omega_{\rm K-He}}$. We point out that at the compensation point, the higher order corrections in $1/\delta\omega$ can become important. While highly dependent on the precise details, one often finds the two eigenvalues' real values to be of the same order of magnitude (see Table~\ref{table:times} for the values for Refs.~\cite{VasilakisThesis,BrownThesis,KornackThesis}).

As an example to why the above discussion could be important, consider the case of Ref.~\cite{BrownThesis}. In Ref.~\cite{BrownThesis}, the detector changes its direction every few seconds. Sudden changes in the magnetic fields are then expected due to possible field penetration as well as  inner thermal noise of the magnetic shields. If the rate with which the system reaches equilibrium after each rotation is slower than the rate of rotations, the system never converges to its steady-state behavior. As a result, the homogeneous terms proportional to the magnetic fields can add significant contributions to the signal. Under realistic laboratory conditions, and even without such a clear intervention in the detector's environmental conditions, sudden changes in the magnetic fields occur, and unless the detector's response-time is fast enough, they can impair the measurements.
Fortunately, since $\Gamma_{\rm slow} \simeq 10\units{sec^{-1}}$, abrupt changes are treated rather efficiently in the systems studied here.

\subsection{The Inhomogeneous Solutions: Response to Oscillatory Fields}

Let us now discuss the steady-state response of the system to high-frequency magnetic fields. The inhomogeneous solutions have three contributions. The first two are from the electron anomalous field and the nuclear anomalous field. These terms relate the alkali's spin measurement 
to those anomalous fields. The third contribution will come from oscillating magnetic fields. It is this that dictates the system's ability to cancel noise at a given frequency of oscillation.

Unlike the homogeneous solutions, whose frequency is determined by the linear system's parameters, an inhomogeneous  term with a certain frequency will only induce an inhomogeneous solution of that frequency. As can be seen from Eq.~\eqref{eq:matrix}, an important change in the presence of an oscillating field with a frequency $\omega$,  is that the steady-state solution is no longer found by taking $\dot{S}_{\rm K}^{\mathbb{C}}=\dot{S}_{\rm He}^{\mathbb{C}}=0$, but rather $\dot{S}_{\rm K}^{\mathbb{C}}=i\omega S_{\rm K}^{\mathbb{C}}$, and $\dot{S}_{\rm He}^{\mathbb{C}}=i\omega S_{\rm He}^{\mathbb{C}}$. Much like in the case of the homogeneous solutions, the actual results are easy to calculate, but have cumbersome formulas. Nonetheless, close to the compensation point, and neglecting $R_{\rm He-K},R_{\rm K-He},R_{\text{He}}$, one finds the approximate closed form solutions,

\bea
\label{eq:highfK}
S^{\mathbb{C}}_{\rm K}(\omega)&=&\frac{P_1(\omega)b_e^{\mathbb{C}}+b_n^{\mathbb{C}}-\frac{\omega \gamma_n B^{\mathbb{C}}}{2\gamma_{\rm He}\lambda \mu_{\rm He} S_{\rm He}^{z(0)}}}{P_2(\omega)},
\\
\label{eq:highfHe}
S_{\rm He}^{\mathbb{C}}(\omega)&=&\frac{\gamma_{\rm He}S_{\rm He}^{z(0)}}{\omega-2 \gamma_{\rm He}\lambda \mu_{\rm He}}\left(\frac{b^{\mathbb{C}}_n}{\gamma_n}+B^{\mathbb{C}}+2\lambda\mu_{\rm K}S_{\rm K}^{\mathbb{C}}\right)\,.~~~~~
\eea
Here $S^{\mathbb{C}}_{\rm K,He}(\omega)$ are the inhomogeneous contributions of the fields $B^{\mathbb{C}}=B^{\mathbb{C}}(\omega)e^{i\omega t},b_n^{\mathbb{C}}=b_n^{\mathbb{C}}(\omega)e^{i\omega t},b_e^{\mathbb{C}}=b_e^{\mathbb{C}}(\omega)e^{i\omega t}$. And $P_1(\omega), P_2(\omega)$ are polynomials of degree one and two, and using the notations of Eq.~\eqref{eq:matrix},
\begin{equation}
P_1(\omega)= -\frac{\gamma_n(\omega+\omega_{\rm He})}{\gamma_e \omega_{\rm He}}\,,
\end{equation}
and
\begin{equation}
P_2(\omega)=\frac{\gamma_n \left((\omega+\omega_{\rm He})(\omega+\omega_K+i\Gamma_K)-\omega_{\rm He-K}\omega_{\rm K-He}\right)}{\gamma_{\rm He}\cdot \omega_{\rm K-He}\cdot S_{\rm He}^{z(0)}}\,.
\end{equation}
Note that due to the ALP field oscillating as $\cos(m_a t+\theta_0)$ with $\theta_0$ an unknown phase, the negative and positive frequencies are mixed, and therefore the final dependence on the ALP field will  be a symmetrized version of Eqs.~\eqref{eq:highfK} and~\eqref{eq:highfHe}.

While it is not yet entirely known what governs the noise spectrum of the comagnetometer at low frequencies (see {\it e.g.} Ref.~\cite{VasilakisThesis,KornackThesis,BrownThesis,katz2019} for calculations of the noise from theoretical arguments, and compare with results from Refs.~\cite{VasilakisThesis,KornackThesis,BrownThesis}), at higher frequencies (usually $\omega \gsim \Gamma_K$ or $|\omega|\gsim |\omega_c|\simeq |\omega_{\rm He}|$) there are reasons to believe that magnetic noise is the dominant factor.  Eq.~\eqref{eq:highfK} shows that such magnetic noises would enter the (measured) alkali's magnetization and thus one can  approximate the $\omega$-dependence  of the signal-to-noise ratio due to suppressed response to ALP-neutron (ALP-electron) interaction, by dividing the coefficient of ${\bf b}_n$ (${\bf b}_e$) with that of ${\bf B}$ in Eq.~\eqref{eq:highfK}. The conclusions is therefore that for ALP-neutron interactions, we  expect an approximately linear decrease in the signal-to-noise sensitivity for high frequencies (the ratio is $\propto 1/\omega$), while  we do not expect such a decrease  for ALP-electron interactions [the ratio is $\propto P_1(\omega)/\omega\sim \mathcal{O}(\omega^0)$].

\section{Effects of Signal Directionality}\label{app:directions}

Here we discuss in detail the procedure for treating signal directionality. For the datasets used in this paper, Refs.~\cite{BrownThesis,VasilakisThesis,KornackThesis}, a simplified treatment sufficed (see Appendix~\ref{app:final}), however here we lay the groundwork for the formal treatment of velocity directionality, which will be relevant in the future with new independent high-resolution data. Throughout this appendix we assume the measurement time $t_{\rm tot}\gg {\rm day}$, as the shortest data-taking session used in our bounds was 4 days long.

The data in Refs.~\cite{BrownThesis,VasilakisThesis} is given in the form of Eq.~\eqref{eq:RomSig} (the data of Ref.~\cite{KornackThesis} is given in a similar, yet not identical form). The directional dependence on the relative ALP velocity is apparent, but the relative velocity of the ALPs with respect to earth is highly model dependent~\cite{Vergados:2016rlh}. Different models can also change the local DM density significantly. Moreover, even for a given model, the local velocity and local density of the ALPs are statistical quantities, and thus need to be treated as such~\cite{Centers:2019dyn}. Our treatment of the effects of the non-deterministic properties of ALPs is presented in Appendix~\ref{app:nonDeterministic}. For the purposes of this appendix, we take the ALPs to  have a constant relative velocity ${\textbf v}$, and a constant density $\rho_{\rm DM}$.

Under these assumptions, we look at the result of plugging the anomalous field implied by the Hamiltonian of Eq.~\eqref{CasperEQ} in the integrand of Eq.~\eqref{eq:RomSig}. We find that
\begin{equation}
\label{Eq:Th}
|A(\omega)|^2=c \left| \int_0^{t_{\rm tot}}dt \cos(E_a t+\theta_0)e^{i\omega t} \left(  \hat{\bf v}\cdot \hat{\bf \sigma}(t)\right) \right|^2,
\end{equation}
where we took the square of the absolute value of the amplitude as we are interested in root mean square over different parameters such as the relative velocity's direction and initial phases. We defined $c\equiv2\rho_{a}|v|^2g_{aNN}^2/(\gamma_n^2 t_{\rm tot})$, in order to make the equations more tractable. $t_{\rm tot}$ is the total measurement time. $E_a$ is the ALP energy, and because the ALP is non-relativistic, $E_a=m_a+m_av^2/2\simeq m_a$ (we will address the importance of deviation from that assumption in Appendix~\ref{app:nonDeterministic}). $\hat{\bf \sigma}$ is the sensitive direction of the detector. We allowed an initial relative phase for the ALP field, $\theta_0$, which we will also discuss in further details in Appendix~\ref{app:nonDeterministic}.

The measurement itself is of the change of polarization in the probe beam behind the cell, rather than ${\bf b}_n\cdot\hat{\bf \sigma}$, and while the change of polarization
is proportional to $S^x_{\rm K}$ (which is proportional to ${\bf b}_n\cdot \hat{\bf \sigma}$), there are calibration factors. These factors are measured individually by the different experiments, with the data given after calibration. The calibration is done by checking low frequency response, and therefore at higher frequencies, a correction is necessary, as was discussed in Appendix~\ref{app:dynamics}. However, for this appendix, we shall assume that the data given is after the necessary additional corrections were made to correct for the higher frequencies -- and thus Eq.~\eqref{Eq:Th} can be assumed as the signal we are given.

To find $\hat{\sigma}$ let us use the coordinate system where $\hat{\bf z}'$ is the direction of earth's rotation axis. We define the $x'-z'$ plane so that at $t=0$, the observer of an experiment on earth is described by $(R_{\oplus}\sin(\theta),0,R_{\oplus}\cos(\theta))$, where $R_{\oplus}$ is the earth's radius, and the observer's latitude coordinate is $\pi/2-\theta$. At time $t$, the observer's position is therefore,
{\small
\begin{eqnarray}
{\bf r}(t)=R_{\oplus}
(\sin(\theta)\cos(\Omega_{\rm SD}t),\sin(\theta)\sin(\Omega_{\rm SD} t),\cos(\theta)),
\end{eqnarray}
}
with $\Omega_{\rm SD}\simeq 2\pi/{\rm day}$ as the sidereal day frequency.

For the experiments of Refs.~\cite{VasilakisThesis,KornackThesis}, the detector's sensitive direction was `up' (the direction of gravity), so that for that case,
\begin{equation}
\label{eq:up}
\begin{aligned}
\hat{\bf \sigma}_1&=
\\
&(\sin(\theta)\cos(\Omega_{\rm SD}t),\sin(\theta)\sin(\Omega_{\rm SD} t),\cos(\theta)).
\end{aligned}
\end{equation}
For the experiment of Ref.~\cite{BrownThesis}, the detector's sensitive direction alternated between the North-South (NS) directions to the East-West (EW) directions every few seconds. Thus, from the second experiment, we have a low-frequency measurement of two different $\hat{\sigma}$,
\begin{equation}
\label{eq:NS}
\begin{aligned}
\hat{\bf \sigma}_2^{\rm NS}&=
\\
&(\cos(\theta)\cos(\Omega_{\rm SD}t),\cos(\theta)\sin(\Omega_{\rm SD} t),-\sin(\theta))\,,
\end{aligned}
\end{equation}
for the NS measurements, and
\begin{equation}
\label{eq:EW}
\hat{\bf \sigma}_2^{\rm EW}=(\sin(\Omega_{\rm SD}t),-\cos(\Omega_{\rm SD} t),0)\,,
\end{equation}
for the EW measurements.

For each of the three directions we plug the appropriate of the three Eqs.~\eqref{eq:up},\eqref{eq:NS},\eqref{eq:EW}, into Eq.~\eqref{Eq:Th}, to get the expected signal. The resulting signal is a complicated function of the many different parameters, and we thus do not show it here. However, as we do want to examine the expected form of the signal, it is useful to look at $\left<\left|A(\omega,\hat{\bf \sigma})\right|^2\right>$, where we would average upon all possible directions $\hat{\bf v}$, and upon the initial angle $\theta_0$. 

For any of the three directions, the resulting averaged signal squared has the form 
\begin{equation}\label{eq:prefactors}
\begin{aligned}
&\left<\left|A(\omega,\hat{\bf \sigma})\right|^2\right>_{\hat{\bf v},\theta_0}=
\\
& c\cdot t_{\rm tot}^2\cdot \sum_{i=1}^6 a_{i}(\hat{\bf \sigma}) \text{sinc}^2((\omega-\omega_i)t_{\rm tot}/2),
\end{aligned}
\end{equation}
where the coefficients $a_i(\hat{\sigma})$ do not depend on the frequencies or the mass, and are in fact only dependent on $\hat{\sigma}(t=0)$. The frequencies $\omega_i$ have six possible values, the sum of one of the two $\{m_a,-m_a\}$ with one of the three $\{\Omega_{\rm SD},-\Omega_{\rm SD},0\}$.  This form is reasonable, as when $t_{\rm tot}\rightarrow\infty$ these terms become delta functions (up to normalization), and as we did not yet include the velocity smearing that shall be discussed in Appendix~\ref{app:nonDeterministic}, the ALPs are indeed infinitely sharp in the frequency range -- albeit possibly shifted due to the earth's rotation.

 As long as $m_a$ is not within $\sim 2\pi/t_{\rm tot}$ of $0$, $\Omega_{\rm SD}$, or $\Omega_{\rm SD}/2$, for all $1\leq i<j\leq 6$, we have
\begin{equation}\label{eq:condition}
|\omega_i-\omega_j|\gg 1/t_{\rm tot}\, .
\end{equation}
When Eq.~\eqref{eq:condition} holds, $A(\omega)$ takes a similar form to Eq.~\eqref{eq:prefactors}, albeit with  $a_{i}(\hat{\bf \sigma})\to  a_{i}(\hat{\bf \sigma},\hat{\bf v})$. When this condition is not met, the signal might smear between different $\omega_i$'s, and take a more complicated form. When specifically $m_{\rm a}\lsim 2\pi/{t_{\rm tot}}$, the effects of $\theta_0$ are not negligible, and its stochastic nature must be accounted for (see Appendix~\ref{app:nonDeterministic} and Appendix~\ref{app:final}). We note that had we not assumed $t_{\rm tot}\gg {\rm day}$, the device's longitude coordinate, and hour at which measurements started would have played a role as well.

\section{Effects of Non-Deterministic Signal}\label{app:nonDeterministic}

Eq.~\eqref{eq:ALPsbfield} presents us with the expected average field for the ALPs throughout the galaxy. However, due to the stochastic nature of the ALP field, $E_a, {\bf v},\theta_0$ and $\rho_a$ should not be treated as their average values throughout the galaxy, when only measured for a short time. Indeed, as we move in the galactic plane, we go through spatial gradients in the ALP field~\cite{Budker:2013hfa}, and the local properties of the ALP field should be thought of as random variables sampled from a distribution centered around the average values. While there is debate in the astrophysics literature as to the size of these gradients, here we take the conservative approach of Ref.~\cite{Guth:2014hsa}, by taking the typical scale of these gradients to be the De-Broglie wavelength of the ALPs, $\sim 2\pi/m_{a}v_{\rm virial}$.

Recently, Ref.~\cite{Centers:2019dyn} has shown how to treat the effect of the stochastic nature of the ALP field, and we base the methods presented in this appendix on theirs. For the case of the ALPs velocity distribution, we also use Ref.~\cite{Lee:2013xxa}. While Ref.~\cite{Lee:2013xxa} was discussing detection of DM scattering via direct detection, their general formulas for finding the relative velocities of virialized DM were useful for our discussion as well.

We will now discuss the different variables that were taken as non-deterministic, and the distributions of these variables. After that, we discuss the coherence time of the signal. The coherence time plays an important role in our treatment of the stochastic nature of the ALP field -- we take an independent sampling of each of the non-deterministic variables every coherence time.

\subsection*{The stochastic nature of the ALP field}

Here we discuss one by one the non-deterministic variables of Eq.~\eqref{eq:ALPsbfield} (identical to the non-deterministic variables that affect ${\bf b}_e$), and what distribution was chosen for them.

{\bf The Initial Signal Phase.}
As we have already briefly mentioned in Appendix~\ref{app:directions}, when $E_a t_{\rm tot}\gsim 2\pi$, the initial phase $\theta_0$ becomes of little importance, as one goes over at least one oscillation of ${\cos}(E_a t+\theta_0)$. Conversely, when $E_a t_{\rm tot} \ll 1$, the signal can be highly dependent on that phase. As this phase is entirely random, we sample it from a uniform distribution between $0$ and $2\pi$.

{\bf The ALP Density.}
The anomalous field (Eq.~\eqref{eq:ALPsbfield}) depends on the square root of the DM energy density. The square root of the ALP energy density, $\sqrt{\rho_a}$, is Rayleigh distributed~\cite{Centers:2019dyn} around $\sqrt{\rho_{\rm DM}}=\sqrt{0.4\ {\rm GeV/cm^3}}$. We therefore sample $\sqrt{\rho_a}$ from the following probability density function,
\begin{equation}
\label{eq:Ray}
p(\sqrt{\rho_a})=\frac{2\sqrt{\rho_a}}{\rho_{\rm DM}}e^{-\rho_a/\rho_{\rm DM}}
\end{equation}

{\bf The Velocity Distribution.} Following Ref.~\cite{Lee:2013xxa}, we take a Standard Halo Model (SHM), where our relative velocity compared to the DM is
\begin{equation}
{\bf v}={\bf v}_{\bigodot}-{\bf v}_{\rm SHM},
\end{equation}
where ${\bf v}_{\bigodot}= (11, 232, 7)\ {\rm km/sec}$, is the velocity of the sun with respect to the galaxy, in galactic coordinates. ${\bf v}_{\rm SHM}$ is the randomly sampled DM velocity in the SHM with respect to the galactic rest-frame. We also use Ref.~\cite{Lee:2013xxa} for their formulas (which we do not reproduce here) to transition Eqs.~\eqref{eq:up},\eqref{eq:EW},\eqref{eq:NS} to the galactic coordinates in which ${\bf v}_{\bigodot}$ is given. 

We have neglected the velocity of the earth with respect to the sun, which would introduce $\sim 10\%$ annual modulation, and we have neglected the velocity from the rotation of the detector around the earth's axis which introduces sub-percent daily modulation. We emphasize that the daily modulations that are discussed in Appendix~\ref{app:directions} are coming from the detector's sensitive-direction's rotation, and not from the small change in the detector's velocity due to earth's rotation. The SHM velocity's probability distribution function is,
\begin{equation}
\label{eq:SHM}
p({\bf v}_{\rm SHM})=\frac{1}{Z}\Theta(v_{\rm esc}-|{\bf v}_{\rm SHM}|)\ e^{-{\bf v}_{\rm SHM}^2/v_{\rm virial}^2},
\end{equation}
where $Z$ is a normalization constant, $\Theta$ is the Heaviside function, $v_{\rm esc}=550\ {\rm km/sec}$ is the galactic escape velocity, $v_{\rm virial}=220\ {\rm km/sec}$ is the virial velocity.

{\bf The Energy Distribution.}
The energy of the non-relativistic ALPs, $E_a=m_a(1+v_a^2/2)$ should be entirely determined by the sampled velocity which was discussed in the previous paragraph. However, as the smearing of the searched frequency introduces a finite coherence time of ALP oscillations, it requires a more thorough discussion, which we perform below.

\subsection*{Effects of finite signal coherence time}

Neglecting the small corrections due to finite galactic escape velocity in the SHM, the spread of velocities gives rise to a coherence time $\tau_a=2\pi/(m_a v^2)\simeq 10^7/m_a$~\cite{Garcon:2017ixh} \footnote{Since the ALP kinetic energy is $\frac{1}{2}m_av_{\rm virial}^2$, some authors use $\tau_a=2\pi/(m_av_{\rm virial}^2/2)$, which is twice as long as the coherence time we use. Our shorter coherence time is conservative, and coincides with Ref.~\cite{Graham:2013gfa} which shows that $\tau_a=2\pi/(m_a v_{\rm virial}^2)$ gives the correct frequency spread from Doppler broadening considerations. Regardless, since the bounds depend only weakly on the exact coherence time, this factor of 2 does not affect the results significantly.}. If a data-taking session is significantly shorter than the coherence time, we assume that the signal is entirely coherent throughout the measurement, i.e. only a single value should be sampled from the distributions discussed in this appendix. 

A coherent signal should scale linearly with $t_{\rm tot}$, the measurement time, while the random noise will scale as $\sqrt{t_{\rm tot}}$, giving rise to SNR [for $S(\omega)$] that scales as $\sqrt{t_{\rm tot}}$. This is why the data of Fig.~\ref{fig:data} is given in the seemingly odd units of $\units{Gauss/\sqrt{Hz}}$. It is therefore expected that even if $t_{\rm tot}$ is increased, the noise spectrum will look the same, while any contribution of the signal will peak over the noise as $t_{\rm tot}$ increases. 

Conversely, if $t_{\rm tot}>\tau_a$, for every $\tau_a$ that passes since the beginning of the measurements, we sample the distributions one more time. Following Ref.~\cite{Budker:2013hfa},  we can also sketch how to understand the dependence of the SNR on $t_{\rm tot}$ after $t_{\rm tot}>\tau_a$. If we assume that $n$ coherence times have passed, $t_{\rm tot}=n\tau_a$, this implies adding incoherently (with random relative phases), $n$ coherent measurements of length $\tau_a$. When $n\gg 1$, the expected measured amplitude of this incoherent summation is approximately the addition in quadratures (as it is a random walk) of the measurements. Therefore, the signal would scale as $\sim \sqrt{n} \tau_a=\sqrt{t_{\rm tot}\cdot \tau_a}$. This would imply that after the coherence time passes, there is no longer an advantage in taking longer measurements (as the signal to noise ratio no longer increases for $t_{\rm tot}>\tau_a$). In a dedicated experiment, as explained in Ref.~\cite{Budker:2013hfa}, and in analogy to the prescribed procedures of Ref.~\cite{Allen:1997ad}, it can be possible to increase sensitivity even after the coherence time passes using curve-fitting of the signal to smeared gaussians. However, since the data analyzed in this paper was not given with sufficient resolution and has gone through several processing procedures, we have not attempted such procedures.

Despite the above discussion, there's an $\mathcal{O}(3)$ improvement in the 95\% C.L. bound in the transition from $t_{\rm tot}\sim \tau_a$ to $t_{\rm tot}\sim 5\tau_a$. This improvement is because when we only have a single sampling of the stochastic distribution, the signal might be unusually small (due to a small $\rho_a$, or a cancellation of ${\bf v}_{\bigodot}$ and ${\bf v}_{\rm SHM}$). Conversely, when $t_{\rm tot}\gsim 5\tau_a$, that probability drops, as we sample 5 different values of our distribution, and while one of them might be small, on average, they would not be consistently small. However as was discussed in the above paragraph, after this improvement at $t_{\rm tot}\sim 5 \tau_a$, the SNR stops improving if one does not use curve fitting. 

In the case of Ref.~\cite{VasilakisThesis}, we are given data that were averaged from multiple measurements. The measurements were taken over a period of $\sim 100$ days. We have assumed that when $\tau_a(m_a)=100$~days, all non-deterministic variables were sampled from a single distribution. However, when $\tau_a\sim 8$~days, most measurements are spaced enough for them to be considered independent samples of the stochastic distribution. We have used a simple interpolation to predict the suppression of the bound due to the stochastic nature of the ALP field, between $\tau_a(m_a)=100$ days and $\tau_a(m_a)=8$ days. Similar, more complicated methods have yielded similar results.

We have used MC simulations for our final bounds and projections presented in Figs.~\ref{fig:gannbounds} and~\ref{fig:gaeebounds}. The procedure of finding the bounds and projections after treating the effects in all other appendices, is described in further details in Appendix~\ref{app:final}.

\section{Effects of Nuclear Structure} \label{app:nuc}

{\bf Nobel Gas Nuclear Structure.}
In this paper we have assumed that the spin of the ${}^3$He is entirely composed of the neutron, as is explained in Ref.~\cite{Stadnik:2014xja}. However, Refs.~\cite{BrownThesis,VasilakisThesis,KornackThesis} claim there is a calibration factor of 0.87 between the neutron spin and the ${}^3$He spin. This $10\%$ modification in the sensitivity to ALP-neutron interactions is not as important as the additional claim for a $\sim 10\%$ contribution of proton spin to the spin of a ${}^3$He nucleus. We have chosen to conservatively assume there is no proton spin in the ${}^3$He, as is claimed in Ref.~\cite{Stadnik:2014xja}. If, however, a proton spin exists, the bounds cast here on the neutrons can be easily converted to bounds on ALP-proton interaction by a simple calibration factor. According to the claims of Refs.~\cite{BrownThesis,VasilakisThesis,KornackThesis}, this factor would translate to a weaker bound on $g_{app}$ by a factor of $\sim$10.

{\bf Alkaline Nucleus.}
We note that the effect of the interaction of a nuclear anomalous field with the alkali's nucleus is negligible. The effect of this interaction is to modify the alkali's energy to have ${\bf b}_e\rightarrow {\bf b}_e+(q-1){\bf b}_n$. Therefore, this correction to the signal from ${\bf b}_n$ will be of order $(q-1)\gamma_{n}/\gamma_e= \mathcal{O}(0.01)$. 
This could allow to look for proton-ALP interactions, if there is a proton spin in the alkali nucleus, but none in the noble nucleus. Note that the bounds will not have the factor of $\gamma_p/\gamma_e\sim 1000$.

\section{Extracting the Bound from the Data}\label{app:final}

Thus far we have described several procedures to treat different effects in Appendices~\ref{app:dynamics},~\ref{app:directions} and~\ref{app:nonDeterministic}. Since not all are relevant for all datasets we use, here we describe the exact procedure we have used to extract the bounds presented in Fig.~\ref{fig:data} and in our future projections.

In all datasets, the direction of the detector is known with respect to earth, so we can use Ref.~\cite{Lee:2013xxa} to find their relative direction with respect to the ALP velocity, sampled from Eq.~\eqref{eq:SHM}. This implies that an ALP would appear as a measurable signal at four frequencies $(\pm (m_a+\Omega_{\rm SD}),\pm (m_a-\Omega_{\rm SD}))$ for the EW measurements of Ref.~\cite{BrownThesis}, and appear as a measurable signal at six frequencies $(\pm (m_a+\Omega_{\rm SD}),\pm (m_a-\Omega_{\rm SD}),\pm m_a)$ for the other 3 datasets we use.

The data in the lower frequencies of Ref.~\cite{KornackThesis} are the only ones studied here with a high enough resolution to allow for a curve fitting procedure to the sinc signal predicted for an ALP (see Eq.~\eqref{eq:prefactors} for a simplified description of the expected curve). However, since the data of Ref.~\cite{KornackThesis} are described as a least square fit and not a Fourier transform, and since their  lowest frequencies induce a far weaker bound than the data of Ref.~\cite{BrownThesis}, we have not attempted to derive the precise shape of an ALP signal with a frequency which is miss-tuned compared to the reciprocal measurement time. Therefore, throughout our analysis, bounds where calculated under the approximation where the squared sinc function is taken to be simply a correctly normalized delta function.

The data of Ref.~\cite{VasilakisThesis}, and the data at higher frequencies from Ref.~\cite{KornackThesis} are given with a much lower resolution than $\Omega_{\rm SD}$. In Ref.~\cite{VasilakisThesis} the data was calculated by smoothing the raw data with a Hann window, while in Ref.~\cite{KornackThesis}, it is only said that the data is the least square fit to an oscillating function. The Hann window smoothed function is as sensitive as a function which has not been smoothed at the points in which the data is given, while the sensitivity is reduced at other points. This means the interpolation of the bound between two data points should not be linear, but dip upwards. Moreover, daily modulations are averaged upon, and the derived limits vary depending on whether the Hann window was used before or after taking the absolute value.
This affects the analysis of the data of Ref.~\cite{VasilakisThesis} and the high frequency data of Ref.~\cite{KornackThesis} (which has been either smoothed or sampled with a resolution that is lower than the natural one). The assumption used in our analysis was that the averaging was done after taking the absolute value. We have also linearly interpolated between the given data points, though as stated here, this interpolation should be treated with caution as depending on the exact processing algorithms used to obtain the data, it may be inaccurate.

Accounting for the effects of the dynamic response of the system is rather simple. A given experiment must calibrate their system in order to know the effect of an anomalous field on the polarization of the probe laser. This is typically done by driving the system away from the compensation point at {\it low frequencies}.  A measured change in polarization is then used to extract a given magnetic field for which the spectral noise density is calculated, as in Eq.~\eqref{eq:RomSig}.  Since at low frequencies, the spin of the alkali is proportional to the anomalous magnetic field [see Eq.~\eqref{eq:romsigCompK}], one may simply divide the spectral noise by the spin value and multiply by the corrected solution at high frequencies, Eq.~\eqref{eq:highfK}.  Recall, that in our complexified spin notations, the real and imaginary values of the alkali spin are both measurable by using probes at different directions.  At low frequencies, $P_2(\omega)$ and $P_1(\omega)/P_2(\omega)$ (corresponding to the alkali's magnetization due to an anomalous electron and neutron magnetic fields respectively; see Eq.~\eqref{eq:highfK}),  are almost purely imaginary, however at higher frequencies they have a real contribution as well, leading to sensitivity for ALPs in the direction parallel to the probe beam. This requires us to specify for the datasets which we analyze at high frequencies the direction of the probe beam -- which for both Refs.~\cite{VasilakisThesis,KornackThesis} was $60^{\circ}$ with the NS directions (and no component in the direction of gravity).

By fitting to a function that is smeared, the effects of the ALP incoherence on the SNR which are described in Appendix~\ref{app:nonDeterministic} become apparent, as different sampled energies would widen the expected signal. However, as no curve fitting was attempted, the effect of finite coherence time has prevented any improvement in SNR after $\tau_a$ passed, as described in the second part of Appendix~\ref{app:nonDeterministic}. For future analysis however, it was assumed that once the coherence time has been reached, the reach still improves as $t_{\rm tot}^{-1/4}$.

At this point, we have a well defined procedure to extract the predicted signal from an ALP.  What remains now is to account for the noise in order to find  the $95\%$ C.L. limit for a given $m_a$, and either $g_{aNN}$ or $g_{aee}$. The problem here is that unlike the more common cases of direct detection experiments, in our setup, noise may in fact theoretically cancel the signal. Therefore, without any model for the noise, a specific measurement could be the remains of a cancellation to an unknown degree between the signal and the noise.

To solve this problem, we need to understand how noise can affect the measurement. Assume that the noise at a given frequency $\omega$, in a given experiment, is with some unknown amplitude $A_{\rm noise}(\omega)$, oscillating with an unknown initial phase. In this case, in order for complete cancellation of the ALP signal at the same frequency, $A(\omega)$, to occur, not only do we need $A(\omega)=A_{\rm noise}(\omega)$, but also for the two phases to exactly match. As these two phases are entirely independent, we expect the relative phase to be a uniformly distributed random variable between $0$ and $2\pi$. Therefore, for a given $A_{\rm noise}(\omega)$, we can easily extract the $95\%$~C.L. of $A(\omega)$ from the measured amplitude at $\omega$. While we have no way of knowing $A_{\rm noise}(\omega)$, we simply take the conservative approach and assume the one that gives the weakest bound. Note that the result is always bounded from below and for any given $A_{\rm noise}(\omega)$, the probability to get  complete cancellation of signal and noise is infinitesimal. 

We find that nearly always, the strongest bounds are found when $A_{\rm noise}(\omega)\to 0$. The main reason for this is the stochastic nature of the signal. The possibility given in the previous paragraph of $A_{\rm noise}(\omega)=A_{\rm signal}(\omega)$ is impossible when the signal is stochastic in nature. Even if the amplitudes are equal for a given ${\bf v},\rho_a$, when we sample the non-deterministic variables, they would not cancel consistently.

The treatment for the dataset of Ref.~\cite{BrownThesis} is a bit more complicated than the other two, and there is more freedom in the choice of statistical test. An ALP of mass $m_a$ would have a measurable amplitude at 5 different points of data, the $|m_a\pm \Omega_{\rm SD}|$ frequencies in both the EW and the NS searches, and the $m_a$ frequency in the NS search. We have taken the mean of these five measurements, though in the future, we expect smarter choices can be made, that could further teach us about the data ({\it e.g.} using the standard deviation of the five measurements to estimate the noise).

We note that when analyzing the data of Ref.~\cite{BrownThesis}, the three masses of $m_a=(0,\frac{1}{2}\Omega_{\rm SD},\Omega_{\rm SD})$ require a special treatment, since for such cases some of the different frequencies at which we attempt to find a signal coincide ({\it e.g.} for $m_a=\frac{1}{2}\Omega_{\rm SD}$, $m_a=-m_a+\Omega_{\rm SD}$), or else we need the zero frequency data which we do not have. The analysis is a simple extension of the previously described procedures, so we do not reproduce it here. 

Before moving to discuss the future projections, we finally note our efficiency estimates. For the data from Ref.~\cite{VasilakisThesis}, we are told that about $35\%$ of the time, the detector was not actively measuring ({\it e.g. } due to the calibration routines), so the effective measurement time is only $65\%$ of the reported $t_{\rm tot}$. As Ref.~\cite{BrownThesis} uses the same procedures, we have taken its efficiency to be $0.65\times 0.5$, as each of the two directions is actively measuring only half the time. For Ref.~\cite{KornackThesis} it is written that the efficiency was between $0.2$ and $0.6$, so we have conservatively taken $0.2$.

Future projections were calculated with a much simpler procedure compared to the bounds, since we cannot be sure of the precise experimental apparatus we will have. The reach is not to be thought of as expected $95\%$~C.L. bound, but as the expected measurable signal. The reach is taken as the sensitivity  described in the text, for a single month of exposure, and under the assumption that the bound when the measurement time is longer than the coherence time improves as $t_{\rm tot}^{1/4}$ (instead of $\sqrt{t_{\rm tot}}$ for $t_{\rm tot}<\tau_a$). To account for the effects of the dynamical response, we assume the sensitivity is weakening linearly for the ${\bf b}_n$ search at $\omega>\omega_c$ (with $\omega_c$ given explicitly in Sec.~\ref{sec:future}).
	
The calculation of the long range spin-dependent interaction bound was written explicitly in Sec.~\ref{sec:future}, and it is effectively a rescaling of the bounds presented by  Ref.~\cite{VasilakisThesis} for their similar experiment. 

\bibliographystyle{JHEP}
\bibliography{DMComag.JHEP}

\end{document}